\documentclass[11pt]{article}
\usepackage{hyperref}
\usepackage{graphicx}
\usepackage{relsize,setspace}  
\usepackage{url}               

\usepackage[dvipsnames]{xcolor}
\usepackage{dcolumn}
\usepackage[ruled,vlined]{algorithm2e}
\usepackage{algorithmic}
\usepackage{xcolor}
\usepackage{graphicx,float}
\usepackage{booktabs}
\usepackage{amsmath,amsfonts,amssymb,mathtools}
\usepackage{makecell}
\usepackage{enumerate}
\usepackage{authblk}
\usepackage{bbm}
\usepackage{apacite}
\usepackage{multirow}
\usepackage{makecell}
\usepackage{cleveref}

\usepackage{booktabs}
\usepackage{siunitx}
\usepackage{fancyhdr}
\usepackage{subfig}
\usepackage{tikz}
\graphicspath{ {plot/} }
\usetikzlibrary{arrows.meta,shapes,shapes.arrows,
shapes.geometric,shapes.multipart,decorations.pathmorphing,
positioning,shapes.swigs,}

\newtheorem{theorem}{Theorem}
\newcommand\numberthis{\addtocounter{equation}{1}\tag{\theequation}}
\newcommand{\indep}{\perp \!\!\! \perp}

\providecommand{\keywords}[1]
{
  \small	
  \textbf{\text{Keywords: }} #1
}

\usepackage{appendix}
\usepackage{titletoc}

\newtheorem{assumption}{Assumption}

\allowdisplaybreaks

\usepackage{float}
\DeclareGraphicsExtensions{.pdf, .jpg, .png}
\setkeys{Gin}{width=0.85\textwidth}

\title{The Multiplicative Quasi-Instrumental Variable Model}
\author[1]{Jiewen Liu
}
\author[2]{Chan Park
}
\author[3]{David Richardson 
}
\author[4,1]{Eric J. Tchetgen Tchetgen
\thanks{
ett@wharton.upenn.edu
Address for correspondence: Eric J. Tchetgen Tchetgen, 407 Academic Research Building, 265 South 37th Street, Philadelphia, PA 19104. Email: ett@wharton.upenn.edu
}}
\affil[1]{Department of Biostatistics, Perelman School of Medicine, University of Pennsylvania}
\affil[2]{Department of Statistics, University of Illinois Urbana-Champaign}
\affil[3]{Joe C. Wen School of Population \& Public Health, University of California, Irvine}
\affil[4]{Department of Statistics and Data Science, Wharton School, University of Pennsylvania}

\newcommand{\qq}{\quad \quad \quad \quad}
\newcommand{\mb}[1]{\mathbb{#1}}
\newcommand{\mc}[1]{\mathcal{#1}}

\newcommand{\ba}[1]{\left( #1 \right)}
\newcommand{\bb}[1]{\left\{ #1 \right\}}
\newcommand{\bc}[1]{\left[ #1 \right]}

\newcommand\norm[1]{{\left\lVert#1\right\rVert}_{P,2}}

\date{}
\begin{document}
\doublespacing
\vspace{-20mm}
\maketitle
\vspace{-8mm}
\begin{abstract}
We introduce the \textit{Multiplicative Quasi-Instrumental Variable (MQIV) model}, a framework for causal inference with unmeasured confounding which leverages an instrument that may be imperfectly exogenous.  We allow the candidate quasi-instrument to have a direct effect on the outcome not mediated by the treatment, thus violating the standard IV exclusion restriction. We establish nonparametric identification of the population average treatment effect on the treated (ATT) under a treatment model that is multiplicative with respect to the quasi-IV and the hidden confounder \cite{hernan2006instruments}. Such a multiplicative treatment model may arise naturally either when treatment occurs only if two independent instrument-driven and confounder-driven causal mechanisms are present; or alternatively, when an instrument’s effect on treatment uptake is inherently heterogeneous and scales with a person's latent propensity, best capturing settings in which it is challenging for a given instrument to overcome a person’s inherent lack of preference for the treatment in view. Importantly, as we establish, the MQIV model is simultaneously agnostic to treatment-effect heterogeneity with respect to hidden confounders, and violation of the core IV exclusion restriction condition. Identification is achieved via a modified Wald ratio estimand, which corrects the bias due to the exclusion restriction violation,  and we propose a new class of estimators that are multiply robust and semiparametric efficient. 
Finally, we evaluate the approach in extensive simulations and an application to evaluate the causal effect of having three or more children on mothers' labor-market engagement.
\end{abstract}

\allowdisplaybreaks
\section{Introduction}
\label{intro}
Instrumental variable methods are a popular approach for making valid inference about the causal effect of a treatment from observational data potentially subject to unmeasured confounding. Informally, an instrumental variable is an observed exogenous factor that is known to induce the treatment of interest and is associated with the outcome only insofar as the induced treatment causally impacts the outcome. Under suitable conditions, a valid instrument can identify a causal effect of interest even when treatment assignment is confounded. The specific causal effect that can be identified with a given instrument typically depends on restrictions on the degree of heterogeneity, either in the effect of the treatment on the outcome model or in the effect of the instrument on the treatment model. Illustrating the former,   \citeA{robins1994correcting} considered an assumption known as ``no current treatment value interaction (NCTI),'' which restricts effect heterogeneity in the outcome model to identify the effect of treatment on the treated. Specifically, \citeA{robins1994correcting} assumed that the average treatment effect among the treated (ATT) is constant across values of the instrument.  Illustrating the latter, in their celebrated paper, \citeA{angrist1996identification} restrict heterogeneity in the treatment model by assuming that the effect of the instrument on treatment is monotone at the individual level. Under this assumption, \citeA{angrist1996identification} formally establish identification of the local average treatment effect (LATE), i.e., the treatment effect for the subpopulation of \emph{compliers}. See also \citeA{permutt1989simultaneous,baker1994paired} for contemporaneous related
developments.   
\\ \\
More recently, \citeA{wang2018bounded} established identification of the population average treatment effect (ATE) under a union of no-heterogeneity conditions: either (i) there is no additive interaction between the treatment and an unmeasured confounder on the mean of the outcome; or (ii) there is no additive interaction between the instrument and an unmeasured confounder in the treatment assignment model; see also \citeA{cui2021necessary} for a related result. \citeA{tchetgen2018marginal} and \shortciteA{michael2024instrumental} extended these results by establishing that assumption (ii) yields IV identification of a marginal causal effect on any scale that would conceivably be identified under unconfoundedness, in both point treatment and time-varying treatment settings.  
\\ \\
Another interesting treatment assignment model takes a familiar multiplicative form, in which the instrument does not interact with the unmeasured confounder on the multiplicative scale for the propensity score. \citeA{hernan2006instruments} considered this model under a certain structural equations model with independent error (SEM-IE) interpretation of the IV model, and showed that it implies the NCTI condition. Therefore, their result shows that under an SEM-IE multiplicative model for the treatment assignment, the ATT is identified by averaging a standard conditional Wald ratio given baseline covariates among the treated. The multiplicative model may arise naturally when treatment occurs only if independent instrument-induced and confounder-induced causal mechanisms are present, or when uptake follows a certain latent index model. More recently, building on the multiplicative model of \citeA{hernan2006instruments},  \citeA{liu2025multiplicative} further showed that under a single world intervention graph (SWIG) interpretation of the IV model encoding a weaker version of latent unconfoundedness, and assuming the multiplicative model holds for a latent propensity score model, the ATT is identified by averaging a certain single arm conditional Wald ratio given baseline covariates among the treated. Also, they showed that the SEM-IE multiplicative model is in fact over-identified for the ATT, and they established the somewhat surprising result that their proposed semiparametric efficient estimator for the ATT under the SWIG multiplicative model remains semiparametric efficient under the more restrictive NPSEM-IE multiplicative model; see \citeA{liu2025multiplicative} for details.  Notably, we emphasize that these results crucially rely on the core exclusion restriction condition that the instrument does not have a direct effect on the outcome other than through the treatment.
\\ \\
A fast-growing body of recent works have aimed to identify a causal effect of a confounded treatment given an imperfect instrument that does not satisfy the exclusion restriction, under an assumption that the additive effect of treatment is constant and a stronger instrument relevance condition, namely that the instrument impacts both the mean and the variance of either the treatment \cite{lewbel2012using, sun2023semiparametric} or the outcome \cite{liu2025quasi}. Other proposed robust methods leverage multiple instruments, assuming some but not necessarily all are valid, without a priori knowledge of which IVs are invalid \cite{kolesar2015identification,kang2016instrumental,bowden2016consistent,guo2018confidence,sun2023semiparametric,kang2025identification}, or they assume that the direct effects of invalid instruments on the outcome are orthogonal to their effects on the treatment, also known as the InSIDE condition (MR-Egger) \cite{bowden2015mendelian}. Crucially, the aforementioned methods a priori rule out the possibility of an additive interaction between the treatment and the hidden confounder in the outcome model, and therefore effectively assume a constant treatment effect.
\\ \\
In this paper, we propose the \textit{Multiplicative Quasi-Instrumental Variable} (MQIV) framework, which identifies the effect of a confounded treatment under the multiplicative model of \citeA{hernan2006instruments} and \cite{liu2025quasi}; however, unlike these previous papers, we establish the somewhat surprising result that the ATT remains nonparametrically identified even if the exclusion restriction assumption is violated due to the presence of a direct effect of the instrument on the outcome other than through the treatment.  Under the proposed framework, we appropriately formally refer to the candidate instrument as a quasi-instrument given that it may or may not satisfy the exclusion restriction condition.  
In contrast with other quasi-IV methods, a salient feature of the MQIV model is that it allows for heterogeneous treatment effects with respect to a hidden confounder, a property we view as particularly advantageous. A key contribution of the paper is to show that a modified version of the \emph{Wald ratio} estimand nonparametrically identifies the ATT under the MQIV model. The modification involves a correction term to account for the presence of a direct effect of the instrument on the outcome. For estimation and inference, we characterize the semiparametric efficiency bound for the proposed modified Wald estimand under a nonparametric model for the observed data distribution; and construct a multiply robust semiparametric estimator which, combined with flexible machine learning estimators of nuisance functions via cross-fitting, attains the efficiency bound provided the nuisance functions can be estimated at sufficiently fast convergence rates. We evaluate the approach in an extensive simulation study and an application aiming to estimate the causal effect of having more than two children on mothers' labor-market engagement.
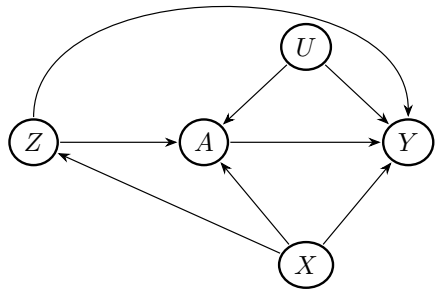
\begin{figure}[b]
\centering  
\scalebox{0.9}{
\begin{tikzpicture}
\tikzset{line width=1pt, outer sep=1pt,
ell/.style={draw,fill=white, inner sep=3pt,
line width=1pt},
swig vsplit={gap=2.5pt, 
inner line width right=0.5pt,
line width right=1.5pt}};

\node[name=A, ell,  shape=ellipse] at (0,0) {$A$};
\node[name=Y, ell,  shape=ellipse] at (3,0) {$Y$};
\node[name=X, ell,  shape=ellipse] at (1.5,-1.8) {$X$}; 
\node[name=U, ell,  shape=ellipse] at (1.5,1.4) {$U$}; 
\node[name=Z, ell,  shape=ellipse] at (-2.5,0) {$Z$}; 

\begin{scope}[>={Stealth[black]},
      every edge/.style={draw=black,line width=0.5pt}]  
\path [->] (X) edge (A);  
\path [->] (X) edge (Y);    
\path [->] (A) edge (Y);     
\path [->] (X) edge (Z);    
\path [->] (Z) edge (A);  
\path [->] (Z) edge [bend left=90] (Y);  
\path [->] (U) edge (A);    
\path [->] (U) edge (Y);  
\end{scope}

\end{tikzpicture}}
\caption{A Graphical Representation of the MQIV model under Assumptions \ref{as:iv1} -- \ref{as:iv3}.}
\label{fig:D1}
\end{figure}

\section{Framework and Identification}
\label{sec:iden}
\subsection{Causal Structure and Notation}
\label{sec:ns}
\noindent
Suppose we observe an i.i.d.\ sample of size $N$ 
consisting of variables \(O=(Y,A,Z,X)\), where \(Y\) is an outcome of interest, \((A,Z)\in\{0,1\}^2\) denote a binary treatment and a binary QIV, and \(X\in\mc{X}\subseteq\mb{R}^d\) is a vector of measured covariates. Let \(U\) denote unmeasured confounders of the causal effect of \(A\) on \(Y\). For each \((a,z)\in\{0,1\}^2\), let \(Y^{a,z}\) be the potential outcome under a hypothetical intervention that sets \((A,Z)\) to \((a,z)\), and let \(A^{z}\) be the potential treatment if, possibly contrary to the fact, \(Z\) were set to \(z\); we also write \(Y^{a}\) for the potential outcome under an intervention that sets \(A\) to \(a\) regardless of $Z$. Our primary target is the marginal average treatment effect on the treated (ATT),
\begin{align*}
\psi := E\ba{Y^{a=1}-Y^{a=0} | A=1}.
\end{align*}
We will first consider the conditional ATT defined by $\psi(X)$ and subsequently recover the marginal ATT denoted by $\psi$ through marginalizing over the covariate distribution among the treated. Throughout, we assume consistency, $Y=Y^{A,Z}$ and $A=A^{Z}$ almost surely. We further assume positivity: $0< Pr(Z=1| X)<1$ and $0< Pr(A=1| Z,U,X)<1$ almost surely. Finally, \(\mb{I}(C)\) denotes the indicator of event \(C\), and \(W \indep H| R\) denotes conditional independence between random variables \(W\) and \(H\) given \(R\).
\\ \\
Throughout, we will consider the following formalization of key IV assumptions:
\begin{assumption}[Relevance]\label{as:iv1} $A \not\indep Z|X$;\end{assumption}
\begin{assumption}[Independence]\label{as:iv2} $Z\indep U,A^{z},Y^{a,z}|X$ ; \end{assumption}
\begin{assumption}[Latent Exchangeability] \label{as:iv3} $A^{z^*}  \indep Y^{a,z}|Z=z,U,X$ for $a\in\bb{0,1}$ and $(z,z)^* \in \bb{0,1}^2$. \end{assumption}
The first assumption $\ref{as:iv1}$ states that $Z$ must be relevant for the treatment within levels of $X$; the second assumption $\ref{as:iv2}$ states that $Z$ and ($U,A^{z},Y^{a,z})$ must be independent in the population conditional on $X$, such that the causal effect of $Z$ on $Y$ and $A$ is unconfounded; while the third assumption $\ref{as:iv3}$ essentially states that $(X,U)$ suffices to account for confounding of the causal effect of $A$ on $Y$, under an NPSEM-IE interpretation of conditional (latent) ignorability; also see  \citeA{hernan2006instruments} who rely on essentially the same condition.
\\ \\
Figure \ref{fig:D1} presents graphical representations of the IV model under assumptions~\ref{as:iv1}--\ref{as:iv3} using a directed acyclic graph (DAG). The arrow $Z \to Y$ encodes the possibility of a direct effect of $Z$ on $Y$ not mediated by $A$, even after accounting for all confounding by conditioning on $(U,X,A)$, which is accommodated by assumption~\ref{as:iv3}. 
\\ \\
The following definitions are used throughout the paper:
\begin{align*}
    &e_z(X) := E\ba{Y|Z=z,X},
    &&e_{a,z}(X) := E\ba{Y|A=a,Z=z,X}, \\
    &\pi_z(X) := f(Z=z|X), 
    && p_z(X) := Pr(A=1 | Z=z, X), \\
    & \rho(X):=Pr(A=1|X),
    && \phi(X):= e_{11}(X) - e_{10}(X), \\
    & \delta^*(X):= \delta(X) - \phi(X) /\bb{p_1(X)-p_0(X)},
    &&\delta^* := E\bb{A\delta^*(X)/Pr(A=1)}\\
    &\delta(X) := \bb{e_1(X) - e_0(X)}/\bb{p_1(X)-p_0(X)} 
    && w(X):= E\bb{Y-A\delta^*(X)-Z\phi(X)|X}   .    
\end{align*}
Moreover, for a sequence of random variables \(\{V_N\}\), we define:
\(\mb{P}(V) = \sum_{i=1}^N V_i/N\), the empirical mean of the observed data;  
\(|\mc{I}_k|\), the cardinality of the set \(\mc{I}_k\), and
\(\mb{P}_{\mc{I}_k}(V) = \sum_{i\in \mc{I}_k} V_i /|\mc{I}_k|\), the empirical mean over the subset \(\mc{I}_k\).  
Furthermore, we use the standard notation \(V_N = O_P(r_N)\) to denote that \(V_N / r_N\) is stochastically bounded , \(V_N = o_P(r_N)\) to denote that \(V_N / r_N\) converges to 0 as \(N \to \infty\), and $\longrightarrow_D$ to denote the convergence in distribution.

\subsection{The QIV \& Data-Generating Mechanisms}
We define $Z$ to be a valid QIV if it satisfies assumptions \ref{as:iv1}--\ref{as:iv3} and the following assumption.
\begin{assumption}[Stable Direct Effect] \label{as:sde} $E\ba{Y^{z=1}-Y^{z=0}|A,Z,U,X}=\beta_Z(X)$ almost surely; for some unrestricted function $\beta_Z(X)$.  \end{assumption}
Assumptions \ref{as:iv3} and \ref{as:sde} allow for a direct effect of the QIV on the outcome and therefore entail a significant relaxation of the standard IV exclusion restriction, which requires
\begin{align*}
\beta_Z(X)=0, \text{almost surely}. \tag{ER} \label{ER}
\end{align*}
Note that \eqref{ER} implies assumption \ref{as:sde}, but not conversely; thus, while our proposed framework accommodates a valid instrument as a special case, it does not require one. Notably, the assumption allows for unrestricted effect heterogeneity of the (additive) direct effect of the instrument; however, it rules out any heterogeneity with respect to either the treatment or the hidden confounder upon conditioning on measured covariate $X$. To make these assumptions more concrete, consider the following nonparametric structural equation model for the mean of the potential outcome $Y^{a,z}$ across values of $(A,Z,X,U)$ for $(a,z)\in\bb{0,1}^2$, compatible with assumptions \ref{as:iv1}--\ref{as:iv2},
\begin{align*}
	E\ba{Y^{a,z}| U,X}
	= \beta_A(U,z,X)a + \beta_U(U,X)+\beta_Z(U,X)z + \beta_X(X),
	\tag{SME}\label{SME}
\end{align*}
where the functions $\beta_A,\beta_U,\beta_Z$ and  $\beta_X$ are a priori unrestricted, and therefore the model is fully saturated. Then, we have that assumption \ref{as:sde} is implied by conditions $\beta_Z(U,X)=\beta_Z(X)$, so that it does not depend on $U$, and $\beta_A(U,z,X)=\beta_A(U,X)$, so that it does not depend on $z$, therefore ruling out any heterogeneity in the direct effect of $z$ on $Y$ encoding a violation of the exclusion restriction, with respect to $(U,A)$.  While this assumption might seem restrictive, it is notable that it is considerably less restrictive than the standard exclusion restriction assumption routinely made under the standard IV model, i.e. $\beta_Z(X)=0$, almost surely.  \\ \\
It is well known that a valid IV satisfying assumptions \ref{as:iv1}--\ref{as:iv3} and \eqref{ER} yields a test of the sharp null of no individual-level treatment effect, and can deliver bounds for various causal estimands \cite{swanson2018partial}. However, without an additional restriction, a valid IV generally does not point identify causal effects such as the ATT. The lack of point identification logically extends to the QIV setting, which means the weaker assumptions \ref{as:iv1}--\ref{as:sde} alone are insufficient for point identification and thus that an additional assumption is required for identification. To make progress, we will consider the multiplicative model for the treatment assignment introduced in \citeA{hernan2006instruments} and \citeA{liu2025multiplicative}. The model rules out a multiplicative interaction between the QIV $Z$ and the hidden confounder $U$ in the propensity score.
\begin{assumption}[Multiplicative Treatment Model] \label{as:miv} $Pr(A=1| Z,X,U)=\exp\bb{\alpha_1(Z,X)+\alpha_2(U,X)}$ where $\alpha_1(\cdot,\cdot)$ and $\alpha_2(\cdot,\cdot)$ are arbitrary functions, with the sole restriction $\alpha_1(0,X)=0$ almost surely, and the natural constraint $Pr(A=1| Z,X,U)\in(0,1)$ almost surely.
\end{assumption}
To build intuition for the multiplicative model, we consider two alternative treatment selection mechanisms that imply the multiplicative model
; throughout we suppress $X$ to simplify notation.
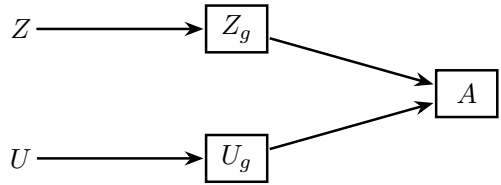
\begin{figure}[b]
\centering
\scalebox{0.95}{
\begin{tikzpicture}[
    >=Stealth,
    line width=1pt,
    outer sep=1pt,
    mech/.style={draw, rectangle, fill=white, minimum width=0.85cm, minimum height=0.65cm, inner sep=2pt},
    var/.style={inner sep=1pt}
]

\node[var] (Z) at (0, 1.8) {$Z$};
\node[var] (U) at (0, 0) {$U$};

\node[mech] (Zg) at (3, 1.8) {$Z_g$};
\node[mech] (Ug) at (3, 0) {$U_g$};

\node[mech] (A) at (6.2, 0.9) {$A$};

\draw[->] (Z) -- (Zg);
\draw[->] (U) -- (Ug);
\draw[->] (Zg) -- (A);
\draw[->] (Ug) -- (A);

\end{tikzpicture}
}
\caption{Graphical representation of the synergy (``AND-gate'') treatment selection model, in which treatment uptake occurs only when both mechanisms $Z_g$ and $U_g$ are present.}
\label{fig:andgate}
\end{figure}
\\ \\
\noindent\textit{\emph{Synergy} (``AND-Gate'') Rule: } Let $Z_g = Z_g(Z,\epsilon_m)\in\{0,1\}$ denote an instrument-driven mechanism depending on $Z$ and an exogenous cause $\epsilon_m$, and let $U_g = U_g(U,\epsilon_t)\in\{0,1\}$ denote a confounder-driven mechanism depending on $U$ and an exogenous cause $\epsilon_t$, with $\epsilon_m \indep \epsilon_t$. A graphical representation of this mechanism is given in \Cref{fig:andgate}. The model states that treatment uptake $A=1$ occurs if and only if both mechanisms are present; otherwise $A=0$: 
\begin{equation}\label{synergy_short}
\mb{I}(A=1)= Z_g\,U_g.
\end{equation}
Although $A$ is a collider of the two mechanisms along the pathway $Z \rightarrow Z_g \rightarrow A \leftarrow U_g \leftarrow U$, conditioning on $\{A=1\}$ under \eqref{synergy_short} implicitly sets $\{Z_g=1,\,U_g=1\}$, which  eliminates any induced association and yields $U \indep Z | A=1$. In contrast, conditioning on $\{A=0\}$ is consistent with three possible values for $(Z_g,U_g)$, namely $(Z_g=1,U_g=0),(Z_g=0,U_g=1)$, or $(Z_g=0,U_g=0)$, which may induce a spurious association $U \not\!\indep Z | A=0$, via collider stratification bias \cite{hernan2004structural}. An immediate consequence of the assumptions $\epsilon_m \indep \epsilon_t$ and \eqref{synergy_short} is therefore the multiplicative treatment selection model
\begin{equation}\label{Multiplicative treatment}
Pr(A=1| U,Z)= g_Z(Z)\,g_U(U),
\end{equation}
obtained by averaging over $(\epsilon_m,\epsilon_t)$ on the right hand side of \eqref{synergy_short}. Importantly, \eqref{Multiplicative treatment} is strictly weaker than the above generative model: independence and synergy imply \eqref{Multiplicative treatment}, but the latter need not imply either condition. Such a synergy (“AND-Gate”) rule has been explored in the sufficient-cause literature on mechanistic interactions. For instance, \citeA{vanderweele2008empirical} illustrate mechanistic interaction of this sort with several examples in which an outcome occurs only when two causes are jointly present, such as “two poisons” acting together. \citeA{vanderweele2010marginal} further apply this idea to arsenic in well water and tobacco smoking for premalignant skin lesions, emphasizing pathways that require both exposures. More generally, multiplicative selection patterns may also arise in broader multi-causal or multi-stage models, for example when distinct causes govern different transitions along a sequential pathway to treatment uptake. Related ideas also appear in the epidemiologic literature on multiplicative disease-risk models; see, for example, \citeA{walter1978additive}, who discuss the ``no-hit'' model under which multiplicative patterns can arise from mechanistic structures involving multiple causal components.
\\ \\
\noindent\textit{Latent Index Model:}
An alternative treatment uptake generative model for the multiplicative model is the following generalized latent index model (GLIM).
Specifically, \citeA{liu2025multiplicative} show that the potential treatment can be generated as
\begin{align}
&A^{z}=\mb{I}\bb{g_Z(z)\,g_U(U)\ge \epsilon_z}, \qquad \epsilon_z\sim Unif(0,1), 	\label{eq:lim-miv}\\
&\textit{for some functions $g_Z(\cdot)$ and $g_U(\cdot)$ such that $g_Z(z)\,g_U(U)\in(0,1)$.} \notag
\end{align}
The GLIM in \eqref{eq:lim-miv} describes treatment uptake as a latent utility index $g_Z(z)\,g_U(U)$ crossing a uniformly distributed threshold $\epsilon_z$.
This representation gives a useful interpretation of the multiplicative treatment model.
The factor $g_U(U)$ may be viewed as capturing a person's latent propensity, readiness, or preference for the treatment, while $g_Z(z)$ captures the instrument-driven multiplicative shift in this latent propensity.
Thus, the effect of the instrument on treatment uptake is inherently heterogeneous: the same value of $Z$ produces a larger absolute change in the latent index for individuals with higher baseline propensity $g_U(U)$, and a smaller change for individuals with low latent propensity.
This is especially natural in settings where the instrument may encourage or facilitate treatment uptake, but may be insufficient to overcome a person's inherent lack of preference, motivation, readiness, or perceived need for the treatment in view. Equivalently, because $g_Z(z)g_U(U)\in(0,1)$ and $\epsilon_z\in(0,1)$, the latent index model can be written on the logarithmic scale as
\[
A^{z}
=
\mb{I}\bb{
\log g_Z(z)+\log g_U(U)\ge \log \epsilon_z
}.
\]
On this scale, the multiplicative model corresponds to an additive latent index with an instrument component $\log g_Z(z)$ and a latent-propensity component $\log g_U(U)$.
However, the induced change in the original probability scale remains multiplicative and therefore depends on the individual's latent propensity. For comparison, \citeA{liu2025multiplicative} also derive GLIMs that correspond to alternative ``no-interaction'' assumptions in the treatment selection model, and contrast these with the GLIM formulation of the monotonicity condition due to \cite{vytlacil2002independence}.
They make the important observation that while the GLIM equivalent representation of monotonicity restricts the degree of individual-level heterogeneity by imposing $\epsilon_z$ to be degenerate, consequently imposing a degenerate joint distribution for the potential treatments $(A^{z=0},A^{z=1})$, the GLIM multiplicative generative model readily incorporates such heterogeneity.
We refer the reader to \citeA{liu2025multiplicative} for a detailed discussion.
\subsection{Comparison with Related IV Models}
\label{sec:model_comparison}
We refer to the model under Assumptions \ref{as:iv1}--\ref{as:miv} as the MQIV model, since it combines the quasi-IV conditions with the multiplicative treatment model. It is useful to compare MQIV with several related IV models for point identification with a binary treatment and a binary instrument. These models differ in whether identification is obtained by restricting heterogeneity in the outcome model, restricting heterogeneity in the treatment-selection model, or maintaining the exclusion restriction.
\\ \\
\noindent\textbf{Robins' no current treatment interaction condition.}
For identification of the ATT, \citeA{robins1994correcting} considered a no-treatment-effect-heterogeneity condition known as no current treatment interaction (NCTI). In the present notation, a version of the Robins IV conditions can be written as
\begin{align}
  E\ba{Y^{a=0}|Z,X} &\indep Z|X, \label{eq:robins_mean_ind}\\
  E\ba{Y^{a=1}-Y^{a=0}|A=1,Z,X} &\indep Z|A=1,X . \label{eq:ncti}
\end{align}
The NCTI condition \eqref{eq:ncti} states that the conditional ATT given $(Z,X)$ is invariant across levels of the instrument among the treated. Under IV relevance and \eqref{eq:robins_mean_ind}--\eqref{eq:ncti}, the conditional ATT is identified by the standard conditional Wald ratio $\delta(X)$. This condition is attractive because it is automatically satisfied under the conditional ATT null and is therefore locally robust for testing the ATT null. However, it restricts the causal effect of primary interest to be constant across strata of the instrument, which may be undesirable when treatment-effect heterogeneity with respect to $Z$ is scientifically plausible.
\\ \\
\noindent\textbf{The Hern\'{a}n--Robins multiplicative IV model under NPSEM-IE.}
\citeA{hernan2006instruments} related NCTI to a multiplicative treatment model under a nonparametric structural equation model with independent errors (NPSEM-IE). In the notation of the present paper, the corresponding formulation imposes a strong latent ignorability and exclusion restriction:
\begin{equation}
  Y^a\indep (A,Z)|X,U,
  \qquad a\in\{0,1\}.
  \label{eq:strong_exclu}
\end{equation}
They prove that, together with IV relevance, IV independence, and the multiplicative treatment model, condition \eqref{eq:strong_exclu} implies NCTI and identifies the ATT by the standard Wald ratio. Thus, the Hern\'{a}n--Robins model does not directly impose a constant treatment effect with respect to the hidden confounder $U$; instead, NCTI follows from the stronger NPSEM-IE-style restrictions together with multiplicative treatment selection.
\\ \\
\noindent\textbf{The MIV model of \citeA{liu2025multiplicative}.}
\citeA{liu2025multiplicative} considered a weaker version of the Hern\'{a}n--Robins formulation by replacing \eqref{eq:strong_exclu} with a single-world condition involving only the treatment-free potential outcome:
\begin{equation}
  Y^{a=0}\indep (A,Z)|X,U .
  \label{eq:weak_exclu}
\end{equation}
Under the remaining valid-IV and multiplicative treatment-model assumptions, this weaker MIV model identifies the treatment-free counterfactual mean among the treated using a single-arm Wald ratio based on the untreated outcome $(1-A)Y$. Consequently, the ATT can be identified without imposing restrictions on $Y^{a=1}$ or on the extent to which treatment effects vary with hidden confounders. Compared with the Hern\'{a}n--Robins model, this formulation removes assumptions on treated potential outcomes that are not necessary for ATT identification, and shifts the key identifying structure to the treatment-selection mechanism rather than to the outcome model.
\\ \\
\noindent\textbf{The proposed MQIV model.}
The proposed MQIV model further relaxes the weak exclusion condition in \citeA{liu2025multiplicative}. Instead of requiring the treatment-free potential outcome to be unaffected by the instrument conditional on $(X,U)$, MQIV allows the candidate instrument to have a direct effect on the outcome, provided that this direct effect is stable with respect to $(A,U)$ conditional on $X$, as formalized in Assumption \ref{as:sde}. Thus, MQIV maintains the multiplicative treatment-selection restriction of MIV but relaxes the requirement that the instrument affect the outcome only through treatment, thus delivering a more robust quasi-instrumental variable method. 

\subsection[Identification Result]{The Identification Result under the MQIV Model } We now state a key nonparametric identification result under the MQIV model.
\begin{theorem}
\label{thm1}
Under the MQIV model, $\psi(X)$ is identified from the observed data by the following expression:
	\begin{align*}
		&\psi(X) =  \delta^*(X)= \delta(X) - \phi(X) /\bb{p_1(X)-p_0(X)}.
	\end{align*}
Additionally, the counterfactual mean parameter $\psi$ is identified by $\int_\mc{X}  \delta^*(X) dF(X|A=1)$. 
\end{theorem}

\Cref{thm1} establishes identification of the ATT in the MQIV model via a modified Wald ratio estimand $\delta^*(X)$, without invoking the exclusion restriction. A salient feature of this result is that it avoids any \emph{a priori} restriction on the extent to which the treatment may interact with the hidden confounder $U$ and therefore the extent of latent heterogeneity of the effects of $A$ in determining $Y$. To make this more concrete, let us again consider structural equation model \eqref{SME} compatible with assumptions \ref{as:iv1}--\ref{as:miv}.
The conditional ATT given $X$ is defined as $E\bb{\beta_A(U,X)|A=1,X}$, which clearly accommodates latent effect heterogeneity by $U$ by averaging over it.  Interestingly, compared to the standard Wald ratio $\delta(X)$, the identification formula $\delta^*(X)$ in the \Cref{thm1} involves backing out the contrast $\phi(X)$ from the reduced form in the numerator of $\delta(X)$. This contrast measures the additive association between the QIV and the outcome among the treated, and is formally shown in the proof of the theorem to identify the causal effect of $Z$ on $Y$ in a hypothetical world in which the treatment is set to zero. Specifically, under the assumptions of the \Cref{thm1}, the contrast $\phi(X)=E\ba{Y|A=1,Z=1,X}-E\ba{Y|A=1,Z=0,X}$ identifies the controlled direct effect $E\ba{Y^{a=0,z=1}-Y^{a=0,z=0}|X}=\beta_Z(X)$, and thus captures the magnitude of the departure from the exclusion restriction assumption on the additive scale. Therefore, subtracting this contrast from the reduced form $E\ba{Y|Z=1,X}-E\ba{Y|Z=0,X}$ in the numerator of the standard Wald ratio effectively renders the QIV a valid instrumental variable, thereby debiasing and allowing it to be interpreted causally.
\\ \\
It is further interesting to compare our identification result to that of \citeA{hernan2006instruments} who considered the multiplicative model with a genuine IV satisfying the exclusion restriction. Specifically, suppose that assumptions \ref{as:iv1}, \ref{as:iv2}, \ref{as:miv}, and \eqref{ER} hold. Then, \citeA{hernan2006instruments} showed that the conditional ATT is nonparametrically identified by $\delta(X)$. For completeness, we reproduce their proof in the Appendix.  The result crucially relies on the exclusion restriction \eqref{ER} and does not appear to hold otherwise. In contrast, \Cref{thm1} drops \eqref{ER} by allowing a direct effect of $Z$ on $Y$ through assumptions \ref{as:iv3}--\ref{as:sde}, and achieves identification by explicitly accounting for the direct effect via the adjustment term $\phi(X)$.
\section{A Semiparametric Efficient Estimator}
\label{sec:semi}
\subsection{Semiparametric Efficiency Bound}
In this section, we derive the Efficient Influence Function (EIF) of the identifying functional $\delta^*$, under a nonparametric model denoted $\mc{M}$  that places no restriction on the observed data distribution.  The following theorem states the main result. 
\begin{theorem}
\label{thm2}
\textbf{(1)} The EIF of $\delta^*$ in model $\mc{M}$ is:
    \begin{align*}
        &EIF(O;\delta^*) = \frac{1}{Pr(A=1)} \bc{A\bb{\delta^*(X) - \delta^* } +  
         \theta(O) }, \\
        &where \ \theta(O) := 
        \frac{\rho(X)}{p_1(X)-p_0(X)}\frac{2Z-1}{\pi_Z(X)}
        \bc{
		Y-A\delta^*(X)-Z\phi(X) - w(X)
		- \frac{A}{p_Z(X)} \bb{Y-Z\phi(L) - e_{10}(X)}
        }
		.
    \end{align*}
    Hence, the corresponding semiparametric efficiency bound for $\delta^*$ in model $\mc{M}$ is $Var\{EIF(O;\delta^*)\}$. \\
    \\
    \textbf{(2)} The EIF is a multiply robust moment equation for $\delta^*$ in the sense that it is an unbiased moment equation, i.e. $E\bb{EIF(O;\delta^*)} =0$, under the union of the following three models: 
\begin{align*}
    &\mc{M}_1: \textit{models for }  p_z(X) \textit{ and }  \pi_z(X) \textit{ are evaluated at their true value, for } z\in\{0,1\}; \\ 
    &\mc{M}_2: \textit{models for } \delta^*(X), e_{1z}(X) \textit{ and } w(X)\textit{ are evaluated at their true value, for } z\in\{0,1\};  \\
    &\mc{M}_3: \textit{models for } \delta^*(X), e_{1z}(X) \textit{ and }  \pi_z(X) \textit{ are evaluated at their true value, for } z\in\{0,1\}.  
\end{align*}    
\end{theorem}
The first component of the EIF, \(A\{\delta^*(X)-\delta^*\}/\Pr(A=1)\), may be viewed as the influence function for the marginal parameter \(\delta^*\) if contrary to fact, the nuisance function \(\delta^*(X)\) were known, obtained by rewriting the identifying expression in \Cref{thm1} as a moment function.  The second component, \(\theta(O)/\Pr(A=1)\), serves as an adjustment accounting for the uncertainty due to having to estimate  \(\delta^*(X)\), and for the nuisance functions needed to estimate the latter.  This correction term is key to the multiple-robustness property in part (2) of \Cref{thm2} and, as shown below, facilitates the construction of an estimator for \(\delta^*\) whose bias is at most second order.  As a direct consequence of multiple robustness, writing the diacritic \(\sim\) to denote model misspecification, we have for \(z\in\{0,1\}\):
\begin{align*}
    &\mc{M}_1:
    E\bb{EIF\ba{O;\delta^*,\tilde{e}_{1z}(X),p_z(X),\pi_z(X), \tilde{w}(X), \tilde{\delta}^*(X)}}=0; \\
    &\mc{M}_2: 
    E\bb{EIF\ba{O;\delta^*,e_{1z}(X),\tilde{p}_z(X),\tilde{\pi}_z(X),w(X),\delta^*(X)}}=0; \\
    &\mc{M}_3: E\bb{EIF\ba{O;\delta^*,e_{1z}(X),\tilde{p}_z(X),\pi_z(X),\tilde{w}(X),\delta^*(X)}}=0.
\end{align*} 
\subsection[Estimator]{Semiparametric Efficient Debiased Estimator}
\label{sec:proposed_estimator}
Our estimator is built on the EIF and uses the Double/Debiased Machine Learning (DDML) framework to allow for flexible machine learning estimation of nuisance functions \cite{chernozhukov2018double}. 
In particular, DDML relies on cross-fitting: we partition the sample \(\{O_i: i=1,\ldots,N\}\) into \(K\) disjoint folds \(\{\mc{I}_1,\ldots,\mc{I}_K\}\). 
For each fold \(k\), nuisance functions are estimated on the training sample \(\mc{I}_k^c\) (the complement of \(\mc{I}_k\)), and the target parameter \(\delta^*\) is evaluated on the held-out fold \(\mc{I}_k\). 
The final estimator averages the resulting \(K\) fold-specific estimates. 
Each fold-specific estimate is computed via the following two steps: (i) estimate nuisance functions on \(\mc{I}_k^c\); (ii) plug them into the EIF and target \(\delta^*\) on \(\mc{I}_k\). \\[0.2cm]

\noindent (\textit{Step 1: Estimation of \(e_{1z}(X)\), \(e_{z}(X)\), \(p_z(X)\), and \(\pi_z(X)\)}): 
For a given split \(k\), we fit \(\hat e^{(-k)}_{1z}(X)\), \(\hat e^{(-k)}_{z}(X)\), \(\hat p^{(-k)}_{z}(X)\), and \(\hat \pi^{(-k)}_{z}(X)\) using the training fold \(\mc{I}_k^c\). 
One may use an ensemble learner such as Super Learner, which combines a user-specified library of candidate estimators and returns a convex combination of their fitted values. 
Super Learner enjoys an oracle-type guarantee: under suitable conditions, its risk is asymptotically no worse than that of the best candidate in the library for a chosen loss (e.g., mean squared error) \cite{van2007super}. With these nuisance estimates, we also form plug-in estimators for the remaining nuisance quantities:
\begin{align*}
&
\begin{aligned}
&\hat{\phi}^{(-k)}(X) = \hat{e}_{11}^{(-k)}(X) - \hat{e}_{10}^{(-k)}(X), \\
&\hat{\delta}^{A(-k)}(X) = \hat{p}_1^{(-k)}(X) - \hat{p}_0^{(-k)}(X),
\end{aligned}
\qquad
\begin{aligned}
&\hat{\delta}^{*(-k)}(X) =
\bb{\hat{e}_1^{(-k)}(X) - \hat{e}_0^{(-k)}(X) - \hat{\phi}^{(-k)}(X)}
\big/\hat{\delta}^{A(-k)}(X), \\
&\hat{\rho}^{(-k)}(X) =
\hat{p}_1^{(-k)}(X)\hat{\pi}_1^{(-k)}(X)
+ \hat{p}_0^{(-k)}(X)\hat{\pi}_0^{(-k)}(X),
\end{aligned}
\\
&\hat{w}^{(-k)}(X)
= \hat{e}_1^{(-k)}(X)\hat{p}_1^{(-k)}(X)
+ \hat{e}_0^{(-k)}(X)\hat{p}_0^{(-k)}(X)
- \hat{\rho}^{(-k)}(X)\hat{\delta}^{*(-k)}(X)
- \hat{\pi}_1^{(-k)}(X)\hat{\phi}^{(-k)}(X).
\end{align*}

\noindent (\textit{Step 2: Cross-fitted estimator of \(\delta^*\)}): 
For each \(k=1,\ldots,K\), we compute a fold-specific estimate \(\hat\delta^{*(k)}\) by evaluating the EIF on \(\mc{I}_k\) with nuisance functions obtained from \(\mc{I}_k^c\). 
We then average over folds to obtain the cross-fitted EIF estimator:
\begin{align*}
    &\hat{\delta}^{*EIF}=\frac{1}{K} \sum_{k=1}^K \hat{\delta}^{*(k)}, \numberthis \label{psi_esti} \\
    &\textit{where} \\
    &\hat{\delta}^{*(k)} =  
    \frac{1}{\mb{P}(A)} 
    \mb{P}_{\mc{I}_k}
	\bc{
    A\hat{\delta}^{*(-k)}(X) + 
    \frac{\hat{\rho}^{(-k)}(X)}{\hat{p}^{{(-k)}}_1(X)-\hat{p}^{{(-k)}}_0(X)} \frac{2Z-1}{ \hat{\pi}^{(-k)}_Z(X)}  
    \bc{
    \begin{aligned}
    &Y-A\hat{\delta}^{*(-k)}(X)-Z\hat{\phi}^{(-k)}(X) - \hat{w}^{(-k)}(X) \\
	&	- \frac{A}{\hat{p}^{{(-k)}}_Z(X)} \bb{Y-Z\hat{\phi}^{(-k)}(X) - \hat{e}^{(-k)}_{10}(X)}
    \end{aligned}
    }
	}
    .
\end{align*}

\subsection[Asymptotic Theory]{Asymptotic Theory for the Proposed Estimator}
\label{stat_property}
Let  $||\cdot||_{P,2}$ denote the $\mc{L}^2(P)$-norm for the arm $A=1$ with respect to the true law $P$ that generates the observed data:
\begin{align*}
    r_{\eta,N}^{(-k)} = \| \hat{\eta}^{(-k)} (X) - \eta(X) \|_{P,2}
    =
    \bc{
    \int_{\mc{X}} 
    \bb{\hat{\eta}^{(-k)} (X) - \eta(X) }^2 \, dF(X|A=1)
    }^{1/2}
    \ , 
\end{align*}
for $\eta \in \bb{p_z,\pi_z,e_z,e_{1z},w,\delta^*}$, $z \in \bb{0,1}$ and $k \in \{1,\ldots,K\}$. Suppose the following conditions hold for  estimated nuisance functions for all $k = 1,\ldots,K$: 
\begin{assumption}[Boundedness of the Estimated Nuisance Functions]
\label{as:bounds}

There exist constants $c_1,c_2>0$ such that $\hat{p}_z^{(-k)}(x) \in [c_1,1-c_1]$, $\hat{\pi}^{(-k)}_z(x) \in [c_1,1-c_1]$, $\hat{e}_z^{(-k)}(x)\in [-c_2,c_2]$, $\hat{e}_{1z}^{(-k)}(x)\in [-c_2,c_2]$, $\hat{w}^{(-k)}(x)\in [-c_2,c_2]$ and $\hat{\delta}^{*(-k)}(x)\in [-c_2,c_2]$ for all $z \in \{0,1\}$ and $x \in \mc{X}$.      
\end{assumption}
\begin{assumption}[Consistent Estimation]
\label{as:consis}
$r_{p_z,N}^{(-k)}$, $r_{\pi_z,N}^{(-k)}$, $r_{e_{1z},N}^{(-k)}$, $r_{w,N}^{(-k)}$ and $r_{\delta^*,N}^{(-k)}$ are $o_P(1).$  
\end{assumption}
\begin{assumption}[Cross-Product Rates] \label{as:rates}
\begin{align*} 
    &
    r_{\delta^*,N}^{(-k)} r_{p_z,N}^{(-k)}, \ 
    r_{\delta^*,N}^{(-k)} r_{\pi_z,N}^{(-k)}, \ 
    r_{\phi,N}^{(-k)} r_{p_1,N}^{(-k)}, \
    r_{w,N}^{(-k)} r_{\pi_z,N}^{(-k)}, \
    r_{e_{10},N}^{(-k)} r_{p_z,N}^{(-k)},
    \textit{ and } \        
    r_{e_{10},N}^{(-k)} r_{\pi_z,N}^{(-k)}
    \textit{ are } o_P\ba{N^{-1/2}}     \textit{ for  $z \in \{0,1\}$}. 
\end{align*}
\end{assumption}
Assumption \ref{as:bounds} requires uniform boundedness of the estimators, with \(p_z(X)\) and \(\pi_z(X)\) bounded away from \(0\) and \(1\), and \(e_z(X)\), \(e_{1z}(X)\), \(w(X)\), and \(\delta(X)\) uniformly bounded on \(\mc{X}\). 
Assumption \ref{as:consis} requires the estimated nuisance functions to converge to their corresponding truth as \(N\to\infty\), a property satisfied by many nonparametric machine learning estimators under standard regularity conditions. 
Assumption \ref{as:rates} further requires the relevant cross-product remainder terms to be \(o_P\!\big(N^{-1/2}\big)\). 
\\ \\ 
Importantly, the latter condition allows for rate trade-offs across nuisance estimators. Specifically, if a suitable subset of the five nuisance functions \(p_z(X)\), \(\pi_z(X)\), \(e_{1z}(X)\), \(w(X)\), and \(\delta(X)\) is estimated sufficiently fast (e.g., faster than \(o_P(N^{-1/4})\)), then the remaining nuisance functions may converge more slowly (potentially slower than \(o_P(N^{-1/4})\)), provided that the list of cross-product terms highlighted in \ref{as:rates} satisfy the rate condition. 
In particular, Assumptions \ref{as:consis}--\ref{as:rates} are satisfied when each nuisance function is estimated at rate \(o_P(N^{-1/4})\). 
For the nuisance quantities that can be directly learned from the observed data, namely \(p_z(X)\), \(\pi_z(X)\), \(e_z(X)\), and \(e_{1z}(X)\), such rates are attainable with a broad class of modern nonparametric machine learning methods.  Notably, \(w(X)\) and \(\delta(X)\) are not directly estimable and are currently constructed via plug-in estimators, so their convergence rates inherit those of their plug-in components.
\\ \\
These assumptions are key to deriving the asymptotic distribution of the estimator in \eqref{psi_esti} in \Cref{th:asym}. Finally, we note that they are consistent with the conditions ensuring EIF unbiasedness in \Cref{thm2} under the union model \(\mc{M}_1\cup\mc{M}_2\cup\mc{M}_3\).
\\ \\
The following \Cref{th:asym} formally establishes the asymptotic normality of the estimator \(\hat{\delta}^{*EIF}\) under the stated assumptions.
\begin{theorem}
\label{th:asym}
    Suppose that assumptions \ref{as:iv1} -- \ref{as:rates} hold. Under model $\mc{M}$, the estimator in \eqref{psi_esti} satisfies
    \begin{align*}
        \sqrt{N}\ba{\hat{\delta}^{*EIF}-\delta^*} \longrightarrow_D N(0,\sigma^2) \  as \  N \rightarrow \infty
        \numberthis \label{asymp-converge}.
    \end{align*}
The variance $\sigma^2$ matches the semiparametric efficient bound for $\delta^*$ in the nonparametric model for the observed data, i.e. $\sigma^2=var\{EIF(O;\delta^*)\}$.
A consistent estimator of $\sigma^2$ is given by
\begin{align*}
        &\hat{\sigma}^2 = \frac{1}{K}\sum_{i=1}^K \mb{P}_{\mc{I}_k}
        \bc{
        \bb{
        \frac{\hat{\gamma}^{(-k)}(O) - A\hat{\delta}^{*EIF}}{\mb{P}(A)}
        }^2
        }
        \\ 
        &where 
        \\
    &\hat{\gamma}^{(-k)}(O)=    
    A\hat{\delta}^{*(-k)}(X) + 
    \frac{\hat{\rho}^{(-k)}(X)}{\hat{p}^{{(-k)}}_1(X)-\hat{p}^{{(-k)}}_0(X)} \frac{2Z-1}{ \hat{\pi}^{(-k)}_Z(X)}  
    \bc{
    \begin{aligned}
    &Y-A\hat{\delta}^{*(-k)}(X)-Z\hat{\phi}^{(-k)}(X) - \hat{w}^{(-k)}(X) \\
	&	- \frac{A}{\hat{p}^{{(-k)}}_Z(X)} \bb{Y-Z\hat{\phi}^{(-k)}(X) - \hat{e}^{(-k)}_{10}(X)}
    \end{aligned}
    }.
    \end{align*}
\end{theorem}
Of note, assumptions \ref{as:consis} and \ref{as:rates} ensure that the scaled (by $\sqrt{N}$) second-order bias of the proposed estimator is $o_P(1)$. Let $z_{1-\alpha/2}$ denote the $(1-\alpha/2)$ quantile of the standard normal distribution. By \Cref{th:asym}, an asymptotic $100(1-\alpha)\%$ Wald-type confidence interval for $\delta^*$ is given by 
\[
\left(\hat{\delta}^{*EIF} - z_{1-\alpha/2} \frac{\hat{\sigma}}{\sqrt{N}}, \; \hat{\delta}^{*EIF} + z_{1-\alpha/2} \frac{\hat{\sigma}}{\sqrt{N}}\right).
\]
\section{Simulation Studies}
\label{sec:sim}
In this section, we evaluate the finite-sample performance of the proposed estimators.  Notably, as discussed earlier, the estimator developed under the MQIV framework does not rely on the exclusion restriction encoded in \eqref{ER} and remains consistent under our assumptions, even when \eqref{ER} fails.  We therefore evaluate its performance in both settings where \eqref{ER} holds and when it does not. 
\\ \\
We generate i.i.d.\ samples $\{Y_i, A_i, Z_i, X_{1i}, X_{2i}, U_i\}$ that are compatible with assumptions \ref{as:iv1}--\ref{as:miv} under the following data-generating process (DGP). We also generate $Y_i^{ER}$ such that samples $\{Y_i^{ER}, A_i, Z_i, X_{1i}, X_{2i}, U_i\}$ satisfy assumptions \ref{as:iv1}, \ref{as:iv2}, \ref{as:miv} and \eqref{ER}. Here, $Y$ reflects a direct effect of $Z$, whereas $Y^{ER}$ is generated under the restriction \eqref{ER} and thus rules out any direct effect of $Z$ on $Y$.  Aside from this difference, the remaining components of the joint data-generating law are identical across the two settings.
\begin{align*}
	& X_1, X_2, U \sim Uniform(0,1) \\
	& Pr(Z=1|X) = \frac{exp(-1 + X_1 + X_2)}{1+exp(-1 + X_1 + X_2)} \\
	& \alpha_1(Z,X) = Z\ba{X_1/2 + X_2/2 + 0.5}\\
	& \alpha_2(U,X) = {-X_1/2 - X_2/2 - U - 0.5} \\	
	& Pr(A=1|Z,X,U) = exp\bb{\alpha_1(Z,X) + \alpha_2(U,X)} 
    \\
	& \beta_U(U,X) = X_1 + X_2^2 + U \\	
	& \beta_A(U,A,X) = U*X_1 + U*X_2 + U^2 \\	
	& \beta_Z(X) = X_1 + X_2 + X_1*X_2 \\	
	& \beta_X(X) = X_1  \\		
	& E\ba{Y^{A,Z}| U,A,Z,X} = \beta_A(U,A,X)A + \beta_U(U,X) + \beta_Z(X)Z + \beta_X(X) \\
	& \epsilon \sim N(0, 0.5^2) \\	
	& Y = E\ba{Y^{A,Z}|U,X,Z} + \epsilon  \\
	& Y^{ER} = Y - \beta_Z(X)Z
\end{align*}
The marginal ATT $\psi$ under this DGP is 0.679. We consider three classes of estimators for the ATT: (1) the modified Wald ratio-based estimator proposed in this paper, (2) the standard Wald ratio-based estimator proposed by \citeA{levis2024nonparametric}, and (3) the single-arm Wald ratio-based estimator proposed by \citeA{liu2025multiplicative}. Estimators (2) and (3) serve as natural benchmarks, as their consistency relies on the validity of \eqref{ER}. It is worth emphasizing that under \cref{as:iv1}-\ref{as:iv3}, \ref{as:miv}, and \eqref{ER}, both (2) and (3) are consistent, but only (3) attains the semiparametric efficiency bound \cite{liu2025multiplicative}. For each class, we study both a plug-in implementation and an IF-based implementation.
\begin{itemize}
	\item[(1)] \textit{(Plug-In)} $\hat{\delta}^{W1}$: The modified Wald ratio-based estimator $\mb{P} \Big[ \bb{A / \mb{P}(A)} \bb{\hat{e}_1(X)- \hat{e}_0(X) - \hat{\phi}(X) } / \{ \hat{p}_1(X) $ $- \hat{p}_0(X)\} \Big]$, which is a plug-in estimator of the identifying formula in \Cref{thm1}. 
	\item[(1)] \textit{(IF-Based)} $\hat{\delta}^{IF1}$: The IF-based estimator $\hat{\delta}^{*EIF}$ proposed in \Cref{th:asym}. This estimator is consistent and semiparametric efficient in the nonparametric model under assumptions \ref{as:iv1}--\ref{as:miv}.
	\item[(2)] \textit{(Plug-In)} $\hat{\delta}^{W2}$: The plug-in estimator based on the standard Wald ratio $\mb{P} \bc{\bb{A / \mb{P}(A)} \cdot \delta(X) }$, which we do not expect to be consistent if the exclusion restriction condition does not hold; however, it should be consistent if \cref{as:iv1}-\ref{as:iv3}, \ref{as:miv} and \eqref{ER} hold. 	
	\item[(2)] \textit{(IF-Based)} $\hat{\delta}^{IF2}$: The estimator based on the IF of the functional $E\bb{A/Pr(A=1) \cdot \delta(X)}$ \cite{levis2024nonparametric}. The estimator is consistent but potentially inefficient in the  semiparametric model where \cref{as:iv1}-\ref{as:iv3}, \ref{as:miv} and \eqref{ER} hold, but not otherwise; see \cite{liu2025multiplicative} for details.
	\item[(3)] \textit{(Plug-In)} $\hat{\delta}^{W3}$: The plug-in estimator based on the single-arm Wald ratio $\mb{P} \bc{ \bb{A / \mb{P}(A)} \delta^{SW}(Y,X)}
    $ \begin{align*}
	    \delta^{SW}(Y,X):=Y
        +
        \frac{
        E\bb{Y(1-A)|Z=1,X} - E\bb{Y(1-A)|Z=0,X}
        }{
        p_1(X) - p_0(X)
        }.
	\end{align*}
	\item[(3)] \textit{(IF-Based)} $\hat{\delta}^{IF3  }$: The estimator based on the IF of the functional $E\bb{A/Pr(A=1) \cdot \delta^{SW}(Y,X)}$ \cite{liu2025multiplicative}. The estimator is consistent and efficient in the semiparametric model where \cref{as:iv1}-\ref{as:iv3}, \ref{as:miv} and \eqref{ER} hold, but not otherwise. 
\end{itemize}


\begin{figure}[!htp]
\centering
\includegraphics[width=1.0\textwidth]{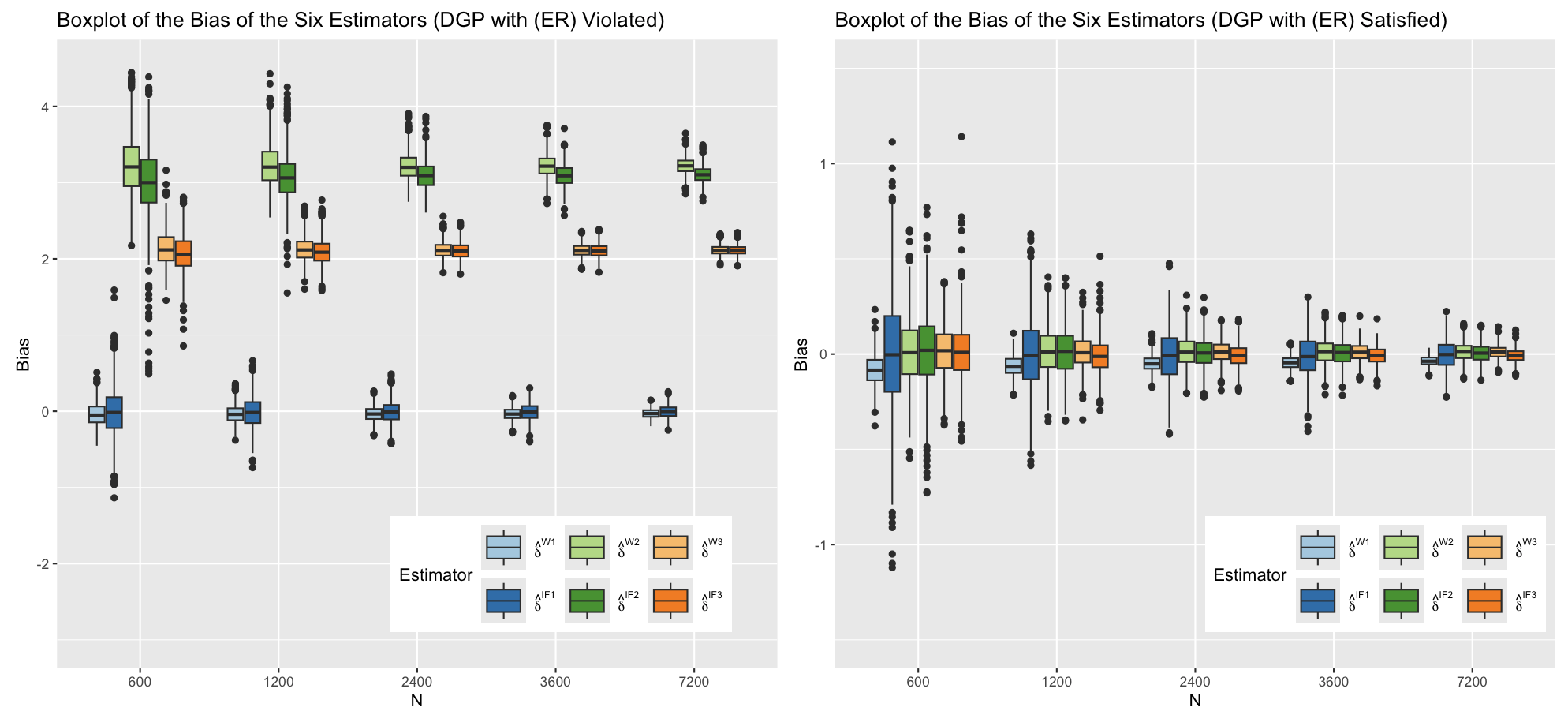}
\caption{
Graphical summary of the simulation results. Each panel shows boxplots of
the bias for the six estimators
\(\hat{\delta}^{W1}\), \(\hat{\delta}^{IF1}\),
\(\hat{\delta}^{W2}\), \(\hat{\delta}^{IF2}\),
\(\hat{\delta}^{W3}\), and \(\hat{\delta}^{IF3}\)
at sample sizes \(N \in \{600,1200,2400,3600,7200\}\).
The left panel visualizes biases under generated data
\(\{Y_i,A_i,Z_i,X_{1i},X_{2i},U_i\}\), where (ER) is violated.
The right panel reports biases under generated data
\(\{Y_i^{ER},A_i,Z_i,X_{1i},X_{2i},U_i\}\), where (ER) holds.
}
\label{fig:sim1}
\end{figure}

\begin{table}[!tb]
\centering
\footnotesize
\setlength{\tabcolsep}{6.5pt} 
\renewcommand{\arraystretch}{1.05}

\begin{tabular}{
l
*{6}{S[table-format=1.3]}
*{6}{S[table-format=1.3]}
}
\toprule
& \multicolumn{12}{c}{\textbf{Estimators}} \\
\cmidrule(lr){2-13}
& \multicolumn{6}{c}{DGP with (ER) Violated}
& \multicolumn{6}{c}{DGP with (ER) Satisfied} \\
\cmidrule(lr){2-7}\cmidrule(lr){8-13}
\textbf{Metrics}
& {$\hat{\delta}^{W1}$} & {$\hat{\delta}^{IF1}$} & {$\hat{\delta}^{W2}$} & {$\hat{\delta}^{IF2}$} & {$\hat{\delta}^{W3}$} & {$\hat{\delta}^{IF3}$}
& {$\hat{\delta}^{W1}$} & {$\hat{\delta}^{IF1}$} & {$\hat{\delta}^{W2}$} & {$\hat{\delta}^{IF2}$} & {$\hat{\delta}^{W3}$} & {$\hat{\delta}^{IF3}$} \\
\midrule

\multicolumn{13}{l}{\textbf{Sample size $N=600$}}\\
Bias     & -0.066 & 0.024 & 3.253 & 2.938 & 2.139 & 2.059 & -0.094 & -0.060 & 0.019 & 0.026 & -0.074 & -0.018 \\
ASE      & 0.154  & 0.343 & 0.415 & 0.471 & 0.232 & 0.259 & 0.080  & 0.291  & 0.177 & 0.195 & 0.129  & 0.144 \\
ESE      & 0.151  & 0.342 & 0.418 & 0.538 & 0.235 & 0.271 & 0.072  & 0.293  & 0.181 & 0.207 & 0.135  & 0.152 \\
Coverage & 0.934  & 0.947 & 0.000 & 0.034 & 0.000 & 0.002 & 0.721  & 0.938  & 0.932 & 0.958 & 0.904  & 0.939 \\
\addlinespace[2pt]

\multicolumn{13}{l}{\textbf{Sample size $N=1200$}}\\
Bias     & -0.052 & 0.022 & 3.235 & 3.045 & 2.129 & 2.090 & -0.090 & -0.049 & 0.021 & 0.026 & -0.058 & -0.021 \\
ASE      & 0.120  & 0.226 & 0.276 & 0.291 & 0.162 & 0.169 & 0.056  & 0.181  & 0.117 & 0.127 & 0.083  & 0.096 \\
ESE      & 0.121  & 0.240 & 0.276 & 0.324 & 0.161 & 0.179 & 0.056  & 0.181  & 0.117 & 0.118 & 0.082  & 0.126 \\
Coverage & 0.930  & 0.944 & 0.000 & 0.002 & 0.000 & 0.002 & 0.574  & 0.952  & 0.912 & 0.962 & 0.882  & 0.946 \\
\addlinespace[2pt]

\multicolumn{13}{l}{\textbf{Sample size $N=2400$}}\\
Bias     & -0.059 & 0.019 & 3.227 & 3.110 & 2.121 & 2.110 & -0.082 & -0.045 & 0.025 & 0.023 & -0.059 & -0.022 \\
ASE      & 0.095  & 0.160 & 0.187 & 0.193 & 0.111 & 0.115 & 0.039  & 0.126  & 0.084 & 0.089 & 0.060  & 0.066 \\
ESE      & 0.096  & 0.141 & 0.188 & 0.190 & 0.110 & 0.114 & 0.039  & 0.125  & 0.085 & 0.086 & 0.060  & 0.059 \\
Coverage & 0.886  & 0.951 & 0.000 & 0.000 & 0.000 & 0.000 & 0.371  & 0.946  & 0.879 & 0.934 & 0.811  & 0.953 \\
\addlinespace[2pt]

\multicolumn{13}{l}{\textbf{Sample size $N=3600$}}\\
Bias     & -0.054 & 0.013 & 3.222 & 3.102 & 2.110 & 2.103 & -0.081 & -0.045 & 0.022 & 0.013 & -0.051 & -0.018 \\
ASE      & 0.087  & 0.132 & 0.155 & 0.156 & 0.092 & 0.093 & 0.032  & 0.106  & 0.071 & 0.073 & 0.050  & 0.055 \\
ESE      & 0.088  & 0.122 & 0.152 & 0.152 & 0.084 & 0.094 & 0.035  & 0.108  & 0.070 & 0.071 & 0.052  & 0.049 \\
Coverage & 0.874  & 0.948 & 0.000 & 0.000 & 0.000 & 0.000 & 0.295  & 0.944  & 0.864 & 0.940 & 0.805  & 0.950 \\
\addlinespace[2pt]

\multicolumn{13}{l}{\textbf{Sample size $N=7200$}}\\
Bias     & -0.046 & 0.007 & 3.222 & 3.109 & 2.115 & 2.114 & -0.075 & -0.038 & 0.019 & 0.009 & -0.031 & -0.012 \\
ASE      & 0.063  & 0.099 & 0.108 & 0.111 & 0.065 & 0.069 & 0.025  & 0.071  & 0.048 & 0.053 & 0.035  & 0.040 \\
ESE      & 0.062  & 0.081 & 0.107 & 0.106 & 0.062 & 0.065 & 0.024  & 0.070  & 0.044 & 0.048 & 0.034  & 0.033 \\
Coverage & 0.869  & 0.954 & 0.000 & 0.000 & 0.000 & 0.000 & 0.135  & 0.945  & 0.861 & 0.951 & 0.807  & 0.948 \\
\bottomrule
\end{tabular}

\caption{Numerical summaries of estimator performance under violation and satisfaction of the exclusion restriction (ER). Each block reports the empirical bias, asymptotic standard error (ASE), empirical standard error (ESE), and empirical coverage of 95\% confidence intervals based on the ASE, across different sample sizes.}
\label{tab:sim_res}
\end{table}

We consider moderate to large sample sizes $N \in \{600,1200,2400,3600,7200\}$ under two regimes: one in which \eqref{ER} holds and one in which it does not. For the IF-based estimator $\hat{\delta}^{IF1}$, we implement the estimation procedure described in \Cref{sec:proposed_estimator}. For the other two IF-based estimators, $\hat{\delta}^{IF2}$ and $\hat{\delta}^{IF3}$, we follow an analogous DDML workflow, using SuperLearner to estimate nuisance functions with cross-validation. Our SuperLearner library includes elastic-net penalized regression, random forests, XGBoost, and generalized additive models as candidate learners \cite{friedman2010regularization,breiman2001random,wood2018mixed,chen2016xgboost}.
\\ \\
Figure~\ref{fig:sim1} summarizes results from 1000 Monte Carlo replications at each sample size. The left panel reports the performance of the six estimators under the DGP in which \eqref{ER} is violated. Across all sample sizes, $\hat{\delta}^{W1}$ exhibits substantially larger bias than $\hat{\delta}^{IF1}$. As expected, $\hat{\delta}^{W2}$, $\hat{\delta}^{IF2}$, $\hat{\delta}^{W3}$, and $\hat{\delta}^{IF3}$ display large, non-diminishing bias because they do not account for the direct effect of $Z$ on $Y$. Compared with $\hat{\delta}^{W2}$ and $\hat{\delta}^{IF2}$, the single-arm estimators $\hat{\delta}^{W3}$ and $\hat{\delta}^{IF3}$ are less biased, since they use the empirical mean to estimate $E\ba{Y^1\mid A=1}$; consequently, violating \eqref{ER} affects only the estimation of $E\ba{Y^0\mid A=1}$.
\\ \\
As $N$ increases, the standard errors of all six estimators decrease. Notably, the empirical coverage of the nominal 95\% confidence interval for $\hat{\delta}^{W1}$ is far below 95\%, whereas the coverage for $\hat{\delta}^{IF1}$ is close to the nominal level, consistent with \Cref{th:asym}. The right panel reports results under the DGP with \eqref{ER} imposed. In this setting, all six estimators are consistent. Moreover, the IF-based estimators $\hat{\delta}^{IF1}$, $\hat{\delta}^{IF2}$, and $\hat{\delta}^{IF3}$ achieve near-nominal 95\% coverage. Among them, $\hat{\delta}^{IF3}$ has the smallest standard error, as expected since $\hat{\delta}^{IF3}$ is based on the efficient influence function. Table~\ref{tab:sim_res} provides the corresponding numerical summary.


\section{Application}
\label{sec:apli}
To illustrate the proposed methods, we analyze the Fertility--Labor Supply Study dataset available in the \texttt{R} package \texttt{ivmte} \cite{shea2023ivmte}. This dataset is based on the study of \citeA{angrist1996children} on the causal impact of childbearing on mother’s weekly working hours. The study sample consists of mothers with at least two children and includes key labor-market outcomes together with baseline covariates used in our analysis, including the mother’s birth year and race/ethnicity indicators. The sample size is $N=$209{,}133. Table~\ref{tab:summary} reports some key descriptive statistics for the study sample.
\\ \\
Our goal is to estimate the average causal effect of an additional child on mother’s weekly working hours, among families ending up with more than two children in the MQIV model proposed in this paper. The causal relationship between fertility and weekly working hours is likely confounded by unmeasured factors such as preferences for market work versus home production, career orientation, and family-related attitudes, which may jointly influence both childbearing decisions and labor-market outcomes. In this setting, IV methods offer a potential way to address unmeasured confounding. In prior work, \citeA{angrist1996children} exploit exogenous variation in fertility induced by (i) the occurrence of twins at the second birth and (ii) the sex composition of the first two children as the candidate IVs for the higher-order birth; in our MQIV setting, we consider the latter and operationalize it as the binary same-sex indicator $Z$. We define $Y$ as weekly hours worked, $A$ as an indicator for having three or more children, $Z$ as an indicator for whether the first two children are of the same sex (i.e., $Z=\mathbb{I}\{\text{the first two children are of the same sex}\}$), and let $X$ collect the baseline covariates. We treat $Z$ as a QIV for $A$ and implement the three IF-based estimators in the Simulation section to assess the effect of an additional child on labor-market outcomes.
\begin{table}[htbp]
\centering
\begin{tabular}{lccccc}
\toprule
 &  & \multicolumn{2}{c}{Exactly two children} & \multicolumn{2}{c}{Three or more children} \\
\cmidrule(lr){3-4}\cmidrule(lr){5-6}
Variable & All & Mixed sex & Same sex & Mixed sex & Same sex \\
\midrule
Sample Size & 209,133 & 72,048 & 67,663 & 31,194 & 38,228 \\
Hours (weekly) & 16.87 (18.37) & 18.48 (18.46) & 18.59 (18.44) & 13.47 (17.85) & 13.53 (17.65) \\
Year of birth & 48.02 (2.81) & 48.21 (2.83) & 48.28 (2.88) & 47.57 (2.65) & 47.59 (2.67) \\
Black & 6.8\% & 6.3\% & 6.4\% & 8.0\% & 7.3\% \\
Hispanic & 2.4\% & 2.0\% & 1.9\% & 3.4\% & 3.3\% \\
Other & 3.4\% & 3.3\% & 3.2\% & 3.6\% & 3.4\% \\
White & 87.4\% & 88.4\% & 88.5\% & 85.0\% & 86.0\% \\
\bottomrule
\end{tabular}
\caption{Descriptive statistics overall and by family size and sex composition of the first two children}
\label{tab:summary}
\end{table}
\\ \\
Within the latent-index representation, assumption~\ref{as:miv} may be plausible in this setting if the instrument $Z$ acts multiplicatively on the latent index governing fertility. This specification allows for latent heterogeneity in fertility responses: the same-sex composition of the first two children may induce a larger shift in the propensity to have an additional child for mothers with stronger preferences for larger families (i.e., with a higher baseline $g(U,X)$), while having a smaller impact for mothers who are less responsive to family-size considerations (low baseline $g(U,X)$).  In contrast to much of the existing literature, we adopt a QIV perspective that allows for a direct effect of $Z$ on $Y$, which mitigates the concern that the same-sex composition of the first two children may directly affect mother’s weekly working hours through channels unrelated to fertility. For example, children of the same sex may share gender-specific resources such as clothing, toys, and extracurricular activities, leading to lower childcare and organizational costs. Moreover, parenting same-sex children may generate ``learning economies,'' whereby accumulated experience with gender-specific needs reduces the time and effort required for caregiving and household logistics. These mechanisms may free up maternal time and increase labor-market participation, inducing a positive direct effect of $Z$ on $Y$. Such a positive $Z \rightarrow Y$ effect may partially offset the negative effect of having an additional child, leading standard IV estimators that ignore the presence of such direct effects to attenuate the estimated $A \rightarrow Y$ effect. To assess the relevance of this concern, we compare $\hat{\delta}^{IF1}$, which accommodates such a potential direct effect of $Z$ on $Y$, to  $\hat{\delta}^{IF2}$ and $\hat{\delta}^{IF3}$, which do not account for a direct effect of $Z$.
\\ \\
We implement the estimation procedure under the DDML framework, adjusting for all available pre-instrument covariates $X$. Nuisance functions are estimated using a stacked SuperLearner that includes elastic-net-penalized regression, random forests, XGBoost and generalized additive models as candidate learners. The results of the three estimators are reported in \Cref{tab:appli}. All estimators yield statistically significant negative treatment effects. The proposed estimator, $\hat{\delta}^{IF1}$, indicates a reduction of approximately 5.3 hours in weekly working hours, whereas the conventional estimators, $\hat{\delta}^{IF2}$ and $\hat{\delta}^{IF3}$, imply a smaller decrease of about 2.6 and 3.5 hours, respectively. This discrepancy between these estimates can be attributed to the uniformly positive $\hat{\phi}(x)$ across values of $x$ in the dataset, suggesting a positive direct effect of the QIV $Z$ on the outcome $Y$ under the MQIV framework. In \Cref{tab:appli}, we also report $\hat{\phi}$, the estimated direct effect of $Z$ on $Y$ averaged over covariates in the treated group, targeting the quantity $E\bb{A/Pr(A=1) \cdot \phi(X)}$.
\begin{table}[!htp]
\centering
\small
\begin{tabular}{@{}p{1.6cm} p{9.2cm} cc@{}}
\toprule
Estimator & Description & Point estimate & 95\% Confidence Interval \\ 
\midrule
$\hat{\delta}^{IF1}$ 
& The estimator based on the IF of the modified Wald ratio proposed in Theorem~\ref{th:asym}. 
& -5.285 & (-8.486, -2.084) \\

$\hat{\delta}^{IF2}$ 
& The estimator based on the IF of the standard Wald ratio. 
& -2.573 & (-5.055, -0.090) \\

$\hat{\delta}^{IF3}$ 
& The estimator based on the IF of the single-arm Wald ratio. 
& -3.500 & (-5.540, -1.460) \\

$\hat{\phi}$ 
& The IF-based estimator of the average direct effect of $Z$ on $Y$ among the treated. 
& 0.134 & (-0.082, 0.349) \\
\bottomrule
\end{tabular}
\caption{Numerical summaries for the data analysis.}
\label{tab:appli}
\end{table}


\section{Discussion}
\label{sec:discuss}
In this paper, we studied identification and estimation of the marginal ATT under the proposed MQIV framework. The framework relaxes the exclusion restriction by allowing a stable direct effect of the QIV on the outcome. We derived the EIF and the semiparametric efficiency bound for the marginal ATT within a nonparametric model for the observed data, and constructed an EIF-based estimator that leverages generic machine learning methods to flexibly estimate the required nuisance functions. We  also formalized sufficient convergence-rate conditions for the nuisance estimators under which the proposed estimator is asymptotically normal. Simulation studies appear to corroborate the theoretical properties of the proposed estimator and additionally compared its performance with the EIF-based estimator of \citeA{liu2025multiplicative}, which is consistent only under the exclusion restriction. Finally, we illustrated the proposed method by estimating the ATT of having more children on mothers' engagement in the labor market using the Fertility–Labor Supply Study dataset.
\\ \\
The framework presented in this paper can be extended in several directions. One interesting question is whether analogous results can be developed for count, binary, or survival outcomes by considering outcome models with appropriate link functions. Such extensions would constitute worthwhile generalizations, and we plan to pursue them in separate work. It would also be of interest to explore applications in environmental epidemiology, where a spatial variable such as location is often used to construct or predict an exposure measure $A$, and interest centers on the association between $A$ and an outcome $Y$. In many such settings, the spatial variable is used only at the exposure-assessment stage and is otherwise ignored in the analysis. Our framework suggests that, if such a variable could instead be viewed as a QIV variable, it may be possible to leverage it to address unmeasured confounding in studies of exposure effects. In addition, the identification results established here rely on key assumptions of the multiplicative model for the treatment and the stable direct-effect condition. Future research may develop sensitivity analysis tools to assess how inference on the ATT might be impacted when these two assumptions do not hold exactly.

\section*{Declaration of the use of generative AI and AI-assisted technologies}
During the preparation of this work the authors used ChatGPT (version 5.4) in order to identify and correct textual errors and improve expression. After using this tool/service the authors reviewed and edited the content as necessary and take full responsibility for the content of the publication.

\bibliographystyle{apacite.bst}
\bibliography{feh}

\clearpage
\setcounter{page}{1}

\markboth{}{}
\thispagestyle{empty}

\begin{center}
{\Large\bf Supplementary Material for}\\[0.5em]
{\Large\bf ``A Multiplicative Quasi-Instrumental Variable Model''}\\[1.5em]

Jiewen Liu, Chan Park, David Richardson, and Eric J. Tchetgen Tchetgen
\end{center}

\vspace{1em}

\begin{abstract}
This supplementary material contains complete proofs of all theoretical
results in the main text, including the identification result, the efficient
influence function, and the asymptotic normality of the proposed estimator.
\end{abstract}

\allowdisplaybreaks

\newpage
\appendix
\begin{appendices}

\section{Proof of the Identification Result}
\subsection{Proof of \Cref{thm1} -- Identification of the ATT}
\label{a:iden}
To show $\psi(X) = \delta^*(X)$, we first note that the mutiplicative model assumption \ref{as:miv} introduces the independence condition $U \indep Z|A=1,X$:
\begin{align*}
	&f(U|A=1,Z,X) 
	\stackrel{Bayes' \  Rule}{=}
	\frac{Pr(A=1|U,Z,X)f(U|Z,X)}{Pr(A=1|Z,X)} \stackrel{A2}{=}\frac{Pr(A=1|U,Z,X)f(U|X)}{Pr(A=1|Z,X)} \\
	&=
	\frac{Pr(A=1|U,Z,X)f(U|X)}{\int_U Pr(A=1|U,Z,X) f(U|X) dU} 
	\stackrel{A4}{=}
	\frac{exp\{\alpha_1(Z,X)+\alpha_2(U,X)\}f(U|X)}{\int exp\{\alpha_1(Z,X)+\alpha_2(U,X)\} dU} 
	=\frac{exp\{\alpha_2(U,X)\}f(U|X)}{\int_U exp\{\alpha_2(U,X)\}f(U|X) dU}  \ \textit{independent of Z.}
\end{align*}
Then, one can show $\beta_Z(X)$ in assumption \ref{as:sde} is identified by $\phi(X)$
\begin{align*}
	&\phi(X)\\
    &=E\ba{Y^{1,1}|A=1,Z=1,X} - E\ba{Y^{1,0}|A=1,Z=0,X} 
    \\
    &\qq  \textit{by independence $U \indep Z|A=1,X$}
    \\
    &= \int_U  \bb{E\ba{Y^{1,1}|U,A=1,Z=1,X} - E\ba{Y^{1,0}|U,A=1,Z=0,X}} dF(U|A=1,X) \\    
	&= \int_U  \bb{
	\begin{aligned}
		&\underbrace{E\ba{Y^{1,1}|U,A=1,Z=1,X} - E\ba{Y^{1,0}|U,A=1,Z=1,X}}_{=\beta_Z(X) \ \textit{ by assumption \ref{as:sde}}} \\
		& + E\ba{Y^{1,0}|U,A=1,Z=1,X}  - E\ba{Y^{1,0}|U,A=1,Z=0,X}
	\end{aligned}
	}
	dF(U|A=1,X) \\   
	&= \beta_Z(X) + \int_U \bb{E\ba{Y^{1,0}|U,A^1=1,Z=1,X}  - E\ba{Y^{1,0}|U,A^0=1,Z=0,X}}
	dF(U|A=1,X) 
    \\
    &\quad \textit{by assumption \ref{as:iv2} that $Z \indep Y^{a,z},A^z,U|X$ and assumption \ref{as:iv3} that $A^{z^*} \indep Y^{a,z}|Z=z^*,U,X$ for all $\ba{z,z}^* \in \bb{0,1}^2$}
    \\	
	&= \beta_Z(X) + \int_U \bb{E\ba{Y^{1,0}|U,X}  - E\ba{Y^{1,0}|U,X}}
	dF(U|A=1,X) \\  
	&= \beta_Z(X).
\end{align*}
Of note, one also has $E\ba{Y|U,A=a,Z=1,X} = E\ba{Y|U,A=a,Z=0,X}+\phi(X)$ for $a=0,1$ because
\begin{align*}
	&E\ba{Y|U,A=a,Z=1,X} = E\ba{Y^{a,1}|U,A^1=a,Z=1,X} = E\ba{Y^{a,1}|U,X} = E\ba{Y^{a,0}|U,X} + \phi(X) \\
	&= E\ba{Y^{a,0}|A^0=a,Z=0,U,X} + \phi(X) = E\ba{Y|A=a,Z=0,U,X} + \phi(X).
\end{align*}
Next, we show $\delta^*(X)$ identifies the conditional ATT
\begin{align*}
	&\delta^*(X) \\
	&= \frac{E\ba{Y|Z=1,X} - E\ba{Y|Z=0,X} - \phi(X)}{Pr(A=1|Z=1,X) - Pr(A=1|Z=0,X)} \\
	&= \frac{E_U\bb{E\ba{Y|U,A,Z=1,X}|Z=1,X} - E\bb{E_U\ba{Y|U,A,Z=0,X}|Z=0,X} - \phi(X)}{Pr(A=1|Z=1,X) - Pr(A=1|Z=0,X)} 
	\\
	&= \frac{
	\bc{
	\begin{aligned}
		& E_U\bb{E\ba{Y|U,A=1,Z=1,X}Pr(A=1|U,Z=1,X) + E\ba{Y|U,A=0,Z=1,X}Pr(A=0|U,Z=1,X)|Z=1,X}  \\
		&- E_U\bb{E\ba{Y|U,A=1,Z=0,X}Pr(A=1|U,Z=0,X) + E\ba{Y|U,A=0,Z=0,X}Pr(A=0|U,Z=0,X)|Z=0,X}  \\
		&-\phi(X)
	\end{aligned}
	}
	}{Pr(A=1|Z=1,X) - Pr(A=1|Z=0,X)} 	
	\\
	&\quad \quad \textit{notice $E\ba{Y|U,A=a,Z=1,X} = E\ba{Y|U,A=a,Z=0,X}+\phi(X)$}
	\\
	&
	= \frac{
	\bc{
	\begin{aligned}
		& E_U\bb{E\ba{Y|U,A=1,Z=0,X}Pr(A=1|U,Z=1,X) + E\ba{Y|U,A=0,Z=0,X}Pr(A=0|U,Z=1,X)|Z=1,X}  \\
		&- E_U\bb{E\ba{Y|U,A=1,Z=0,X}Pr(A=1|U,Z=0,X) + E\ba{Y|U,A=0,Z=0,X}Pr(A=0|U,Z=0,X)|Z=0,X} \\
		&+ \phi(X) E_U\bb{Pr(A=1|U,Z=1,X)  + Pr(A=0|U,Z=1,X)} - \phi(X)
	\end{aligned}
	}
	}{Pr(A=1|Z=1,X) - Pr(A=1|Z=0,X)}
	\numberthis \label{single_world}
	\\
	&\quad \quad \textit{rearrange the numerator}
	\\
	&= \frac{
	\bc{
	\begin{aligned}
		& E\bb{E\ba{Y|U,A=1,Z=0,X}Pr(A=1|U,Z=1,X) - E\ba{Y|U,A=1,Z=0,X}Pr(A=1|U,Z=0,X)|X}  \\
		&+ E\bb{E\ba{Y|U,A=0,Z=0,X}Pr(A=0|U,Z=1,X) - E\ba{Y|U,A=0,Z=0,X}Pr(A=0|U,Z=0,X)|X} 
	\end{aligned}
	}
	}{Pr(A=1|Z=1,X) - Pr(A=1|Z=0,X)} 		 		
	\\
	&\quad \quad \textit{notice $E\ba{Y|U,A=a,Z=z,X}=E\ba{Y^{a,z}|U,A^z=a,Z=z,X} = E\ba{Y^{a,z}|U,X}$} 
\\
	&\quad \quad \textit{by assumption \ref{as:iv2} that $Z\indep Y^{a,z},A^z,U|X$ and assumption \ref{as:iv3} that $A^{z^*}\indep Y^{a,z}|Z=z,U,X$ for $(z,z^*) \in \bb{0,1}^2$}
	\\
	&=
	E_U\bc{ \bb{E\ba{Y^{a=1,z=0}|U,X}-E\ba{Y^{a=0,z=0}|U,X}}  \frac{ Pr(A=1|U,Z=1,X) - Pr(A=1|U,Z=0,X)}{Pr(A=1|Z=1,X) - Pr(A=1|Z=0,X)}|X}
	\\
	&= 
	\int_U E\ba{Y^{a=1,z=0} - Y^{a=0,z=0}|U,X} \frac{f(U|A=1,X)}{f(U|X)} f(U|X)dU 	
    \numberthis \label{bayes}	\\
    &=\int_U E\ba{Y^{a=1,z=0} - Y^{a=0,z=0}|U,X} f(U|A=1,X)dU 	 \\
    &=\int_U E\ba{Y^{a=1,z=0} - Y^{a=0,z=0}|A^{z=0}=1,Z=0,U,X} f(U|A=1,Z=0,X)dU 	 \\    
    &=\int_U E\ba{Y^{a=1,z=0} - Y^{a=0,z=0}|A=1,Z=0,U,X} f(U|A=1,Z=0,X)dU 	 \\   
	&= E\ba{Y^{a=1,z=0}-Y^{a=0,z=0}|A=1,Z=0,X}.
\end{align*}
Equation \eqref{bayes} holds by
\begin{align*}
	&\frac{Pr(A=1|U,Z=1,X) - Pr(A=1|U,Z=0,X)}{Pr(A=1|Z=1,X) - Pr(A=1|Z=0,X)} \stackrel{Bayes' Rule}{=} \frac{\frac{f(U|A=1,Z=1,X)Pr(A=1|Z=1,X)}{f(U|Z=1,X)} -\frac{f(U|A=1,Z=0,X)Pr(A=1|Z=0,X)}{f(U|Z=0,X)}}{Pr(A=1|Z=1,X) - Pr(A=1|Z=0,X)}\\
	&= \frac{\frac{f(U|A=1,X)Pr(A=1|Z=1,X)}{f(U|X)} -\frac{f(U|A=1,X)Pr(A=1|Z=0,X)}{f(U|X)}}{Pr(A=1|Z=1,X) - Pr(A=1|Z=0,X)} =\frac{f(U|A=1,X)}{f(U|X)} \quad \ba{\textit{by $U\indep Z|X$ and $U \indep Z|A=1,X$}}.
\end{align*}
By parallel algebra, one can also show $\delta^*(X)$ identifies  $E\ba{Y^{a=1,z=1}-Y^{a=0,z=1}|A=1,Z=1,X}$ by letting equation \eqref{single_world} in the previous derivation be (difference colored in red)
$$\frac{
	\bc{
	\begin{aligned}
		& E_U\bb{E\ba{Y|U,A=1,Z={\color{red}1},X}Pr(A=1|U,Z=1,X) + E\ba{Y|U,A=0,Z={\color{red}1},X}Pr(A=0|U,Z=1,X)|Z=1,X}  \\
		&- E_U\bb{E\ba{Y|U,A=1,Z={\color{red}1},X}Pr(A=1|U,Z=0,X) + E\ba{Y|U,A=0,Z={\color{red}1},X}Pr(A=0|U,Z=0,X)|Z=0,X} \\
		&+ \phi(X) {\color{red}E_U\bb{Pr(A=1|U,Z=1,X)  + Pr(A=0|U,Z=1,X)}} - \phi(X)
	\end{aligned}
	}
	}{Pr(A=1|Z=1,X) - Pr(A=1|Z=0,X)},$$
and then proceed with the analogous steps.
\\ \\
Finally, we conclude the proof with 
\begin{align*}
    &\delta^*(X)=E\ba{Y^{a=1,z=0}-Y^{a=0,z=0}|A=1,Z=0,X} = E\ba{Y^{a=1,z=1}-Y^{a=0,z=1}|A=1,Z=1,X} \\
    &= E\ba{Y^{a=1}-Y^{a=0}|A=1,X},
\end{align*}{} because 
\begin{align*}
	&\delta^* = E\ba{Y^{a=1,z=0}-Y^{a=0,z=0}|A=1,Z=0,X} = E\ba{Y^{a=1,z=1}-Y^{a=0,z=1}|U,A=1,Z=1,X} \\
	&\textit{by $U\indep Z|A=1,X$} \\
	&E\ba{Y^{a=1,z=0}-Y^{a=0,z=0}|A=1,Z=0,X} = \int_U E\ba{Y^{a=1,z=0} - Y^{a=0,z=0}|U,A=1,Z=0,X} f(U|A=1,Z=0,X) dU \\
	&E\ba{Y^{a=1,z=1}-Y^{a=0,z=1}|A=1,Z=1,X} = \int_U E\ba{Y^{a=1,z=1} - Y^{a=0,z=1}|U,A=1,Z=1,X} f(U|A=1,Z=1,X) dU
	\\
	&\textit{by assumption \ref{as:iv2} $Z\indep Y^{a,z},A^z,U|X$ and assumption \ref{as:iv3} $A^{z^*}\indep Y^{a,z}|Z=z,U,X$ for $(z,z^*) \in \bb{0,1}^2$  and consistency} \\
	&E\ba{Y^{a=1,z=0}-Y^{a=0,z=0}|A=1,Z=0,X} = \int_U E\ba{Y^{a=1} - Y^{a=0}|U,A=1,Z=0,X} f(U|A=1,Z=0,X) dU \\
	&E\ba{Y^{a=1,z=1}-Y^{a=0,z=1}|A=1,Z=1,X} = \int_U E\ba{Y^{a=1} - Y^{a=0}|U,A=1,Z=1,X} f(U|A=1,Z=0,X) dU
	\\
	\implies &\delta^* = Pr(Z=1|A=1,X) \delta^* + Pr(Z=0|A=1,X)\delta^* \\
	&=\int_U E\ba{Y^{a=1} - Y^{a=0}|U,A=1,Z=1,X} Pr(Z=1|A=1,U,X)f(U|A=1,X) dU \\
	&\qq +  \int_U E\ba{Y^{a=1} - Y^{a=0}|U,A=1,Z=0,X} Pr(Z=0|A=1,U,X) f(U|A=1,X) dU \\
	&=\int_U E\ba{Y^{a=1} - Y^{a=0}|U,A=1,X} f(U|A=1,X) dU =  E\ba{Y^{a=1} - Y^{a=0}|A=1,X}.
\end{align*}

\subsection{Identification of the ATT under Assumptions \ref{as:iv1}, \ref{as:iv2}, \ref{as:miv}, and \eqref{ER}}
We prove the conditional ATT is nonparametrically identified by $\delta(X)$ under assumptions \ref{as:iv1}, \ref{as:iv2}, \ref{as:miv}, and \eqref{ER}.
\begin{align*}
    &\delta(X) \\
    &= \frac{E\ba{Y|Z=1,X}-E\ba{Y|Z=0,X}}{Pr(A=1|Z=1,X)-Pr(A=1|Z=0,X)} \\
    &= \frac{E\bb{Y^{1,1}A + Y^{0,1}(1-A)|Z=1,X}-E\bb{Y^{1,0}A + Y^{0,0}(1-A)|Z=0,X}}{Pr(A=1|Z=1,X)-Pr(A=1|Z=0,X)} \\
    &= \frac{\int_U \bc{E\bb{Y^{1,1}A + Y^{0,1}(1-A)|U,Z=1,X}-E\bb{Y^{1,0}A + Y^{0,0}(1-A)|U,Z=0,X}} dF(U|X)}{Pr(A=1|Z=1,X)-Pr(A=1|Z=0,X)} \\ 
    &\quad \textit{by assumption \ref{as:iv2} that $Z \indep Y^{a,z},A^z,U|X$} \\
    &= \frac{\int_U \bc{E\bb{Y^{1,1}A + Y^{0,1}(1-A)|U,X}-E\bb{Y^{1,0}A + Y^{0,0}(1-A)|U,X}} dF(U|X)}{Pr(A=1|Z=1,X)-Pr(A=1|Z=0,X)} \\     
    &= \frac{\int_U \bc{E\bb{ \ba{Y^{1,1} - Y^{1,0}} A |U,X} - E\bb{ \ba{Y^{0,1}-Y^{0,0}} (1-A)|U,X}} dF(U|X)}{Pr(A=1|Z=1,X)-Pr(A=1|Z=0,X)} \\         
    &= \frac{\int_U \bc{E\bb{ Y^{1,1} - Y^{1,0} |A=1,U,X} Pr(A=1|U,Z=1,X) - E\bb{ Y^{0,1}-Y^{0,0}|A=0,U,X} Pr(A=1|U,Z=0,X)} dF(U|X)}{Pr(A=1|Z=1,X)-Pr(A=1|Z=0,X)} \\  
    &\quad \textit{by \eqref{ER} that $E\ba{Y^{z=1}-Y^{z=0}|A,Z,U,X}=0$ almost surely} \\    
    &= \frac{\int_U E\ba{Y^1 - Y^0|U,X} \bb{Pr(A=1|U,Z=1,X) - Pr(A=1|U,Z=0,X)} dF(U|X)}{Pr(A=1|Z=1,X)-Pr(A=1|Z=0,X)} \\    
    &= \frac{\int_U E\ba{Y^1 - Y^0|U,X} \bc{Pr(A=1|U,Z=1,X) - Pr(A=1|U,Z=0,X)}dF(U|X)}{Pr(A=1|Z=1,X)-Pr(A=1|Z=0,X)} \\
    &\int_U E\ba{Y^1 - Y^0|U,X} \frac{f(U|A=1,X)}{f(U|X)} f(U|X) dU \\
    &= E\ba{Y^1 - Y^0|A=1,X}
\end{align*}

\section{Proof of the Efficient Influence Function}
\subsection{Proof of \Cref{thm2}}
\label{a:semi}
In this section, we first prove part \textbf{(1)} and then part \textbf{(2)} of \Cref{thm2}.

\textbf{\textit{Deriving the EIF}}

We assume regular parametric submodels $\mc{P}_v = \bb{\mc{P}_v : v \in \mathbb{R}}$, indexed by a real-valued parameter \(v\), within the nonparametric model \(\mc{M}\) of interest \cite{newey1990semiparametric, tsiatis2006semiparametric}. Let \(P_v\) denote the probability distribution under parameter value \(v\), with \(P = P_{v=0}\) representing the true data-generating distribution. Denote the observed data by \(O = (Y, A, Z, X)\), and let \(f_v\) be the density (or probability mass) function of \(P_v\). We write \(\nabla_v = \frac{\partial}{\partial v}\), and denote the pathwise derivative of \(f_v\) by \(f'_v = \nabla_v f_v\), and likewise for the probability mass function: \(Pr'_v = \nabla_v Pr_v\). The score function for the submodel is given by \(S_v = f'_v / f_v\) (or \(S_v = Pr'_v / Pr_v\) in the discrete case). We omit the subscript $v$ when $v=0$, meaning the quantity is defined under the true law $P$. Let $\mathcal{L}_{2}(W) = \bb{ h(W) | E\bb{h^2(W) }< \infty }$ denote the Hilbert space of square-integrable functions of a random variable $W$, equipped with the inner product $\langle h_1,h_2 \rangle = E\bb{ h_1(W) h_2(W) }$. 

The following properties T1, T2 and T3 are used for the following EIF derivation:
\begin{align*} 
&\text{T1: }  E\bb{h(X) S(Y|X)} =0 
\ , \quad \forall h(X) \in \mc{L}_2(X);
\quad 
\textit{(score has expectation 0 and by iterated expectation;)}\\
&\text{T2: }  E[h(X) \bb{ Y - E(Y|X)}] =0 \ , \quad \forall h(X) \in \mc{L}_2(X);
\quad 
\textit{(by iterated expectation;)} \\
&\text{T3: }  S(O) = S(Y,A,Z,X) = S(Y|A,Z,X) + S(A|Z,X) + S(Z|X) + S(X) \ ; 
\quad 
\textit{(by the chain rule.)}
\end{align*}

We begin deriving the EIF for $\delta^*= E\bb{A\delta^*(X)/Pr(A=1)}$. The canonical gradient for $\delta^*_v = E_v\bb{A\delta^*_v(X)/Pr_v(A=1)}$ in the parametric submodel $\mc{M}_v$ is given by
\begin{align*}
	&\frac{\partial}{\partial v} \delta^*_v 
 \\
	&=\frac{\partial}{\partial v} \int\delta^*_v(X) \frac{A}{Pr_v(A=1)} f_v(A,X) d(A,X)
 \\
	&= \int \frac{\partial}{\partial v} \bb{ \delta^*_v(X) \frac{A}{Pr_v(A=1)} f_v(A,X) } d(A,X)\\
	&= \int \delta^*_v(X) \frac{A}{Pr_v(A=1)} \frac{f'_v(A,X)}{f_v(A,X)} f_v(A,X)  d(A,X) + \int  \bb{\nabla_{v} \delta^*_v(X) }\frac{A}{Pr_v(A=1)} f_v(A,X)  d(A,X) 
 \notag  
 \\ 
	&\quad \quad \quad +\int \delta^*_v(X) \bb{-\frac{A}{Pr_v(A=1)}\frac{Pr'_v(A=1)}{Pr_v(A=1)} } f_v(A,X)  d(A,X) 
 \\
	&= 
	\underbrace{E_v\bb{\delta^*_v(X) \frac{A}{Pr_v(A=1)} S_v(A,X)}}_{\text{Part 1}} + 
	\underbrace{E_v\bb{ (\nabla_{v}\delta^*_v(X)) \frac{A}{Pr_v(A=1)} }}_{\text{Part 2}} - 
	\underbrace{E_v\bb{\delta^*_v(X) \frac{A}{Pr_v(A=1)} S_v(A=1)}}_{\text{Part 3}} 
\end{align*}
The second-to-last equation above holds by the chain rule of differentiation.  We find alternative representations of Parts 1-3 as follows: 

Part 1:

By T1, one has
\begin{align*}
	& E_v\bb{\delta^*_v(X) \frac{A}{Pr_v(A=1)} S_v(A,X)} = 
	E_v\bb{\delta^*_v(X) \frac{A}{Pr_v(A=1)} S_v(Y,A,Z,X)} = 
	E_v\bb{\delta^*_v(X) \frac{A}{Pr_v(A=1)} S_v(O)}.
\end{align*}

Part 3:
\begin{align*}
	&E_v\bb{\delta^*_v(X) \frac{A}{Pr_v(A=1)} S_v(A=1)} \\
	&= \int \delta^*_v(X)  \frac{A}{Pr_v(A=1)} S_v(A=1) f_v(A,X) d(A,X) \\
	&= \int \delta^*_v(X)  \frac{A}{Pr_v(A=1)} S_v(A=1) f_v(A=1,X) d(A,X) \\		
	&= \int \delta^*_v(X)  S_v(A=1) f_v(X|A=1) d(A,X) \\
	&= S_v(A=1) E_{v}\{\delta^*_v(X)|A=1\}\\
	&= S_v(A=1) \delta^*_v\\
	&= E_v\bb{\delta^*_v \frac{A}{Pr_v(A=1)} S_v(A)} \\
	&= E_v\bb{\delta^*_v \frac{A}{Pr_v(A=1)} S_v(Y,A,Z,X)} \\
	&= E_v\bb{\delta^*_v \frac{A}{Pr_v(A=1)} S_v(O)}
\end{align*}
The third-to-last equality follows by noting that $E_v\bb{A/Pr_v(A=1) S_v(A)}=S_v(A=1)$, and the second-to-last equality holds by T1.

Part 2:
\begin{align*}
	&E_v\bc{\bb{\nabla_{v}\delta^*_v(X)} \frac{A}{Pr_v(A=1)} } \\
	&=\int \bb{\nabla_{v}\delta^*_v(X)} \frac{A}{Pr_v(A=1)} f_v(X,A) d(X,A)
\end{align*}
Note that $\nabla_{v}\delta^*_v(X)$ can be expressed as follows by the chain rule:
\begin{align}
	&\nabla_{v}\delta^*_v(X) 
	= \frac{
	\bb{\nabla_{v} \delta_{Yv}(X)} \delta_{Av}(X) - \delta_{Yv}(X) 
	\bb{\nabla_{v}\delta_{Av}(X)} ]
	}
	{\bb{\delta_{Av}(X)}^2},
    \label{eq-delta}
\end{align}
where
\begin{align*}
	&\delta_{Yv}(X) := E_v\ba{Y|Z=1,X} - E_v\ba{Y|Z=0,X} - \bc{E_v\ba{Y|A=1,Z=1,X} - E_v\ba{Y|A=1,Z=0,X}}, \\
	&\delta_{Av}(X) := Pr_v(A=1|Z=1,X)-Pr_v(A=1|Z=0,X).
\end{align*}
Hence, it suffices to find expressions of $\nabla_{v}\delta_{Yv}(X) $
$$\nabla_{v}\delta_{Yv}(X) = \nabla_{v}E_v\ba{Y|Z=1,X} - \nabla_{v}E_v\ba{Y|Z=0,X} - \bc{\nabla_{v}E_v\ba{Y|A=1,Z=1,X} - \nabla_{v}E_v\ba{Y|A=1,Z=0,X}}$$ 
and $\nabla_{v}\delta_{Av}(X)$ 
$$\nabla_{v}\delta_{Av}(X)= \nabla_{v}Pr_v(A-1|Z=1,X) - \nabla_{v}Pr_v(A=1|Z=0,X).$$ We first derive an expression of $\nabla_{v}E_v\ba{Y|Z=z,X}$ and  $\nabla_{v}E_v\ba{Y|A=1,Z=z,X}$:
\begin{align*}
	&\nabla_{v} E_v\ba{Y|Z=z,X} \\
	&= \frac{\partial}{\partial v} \int Yf_v(Y|Z=z,X)dY \\
	&= \int Y\frac{f'_v(Y|Z=z,X)}{f_v(Y|Z=z,X)} f_v(Y|Z=z,X) dy\\ 
	&=E_v\bb{YS_v(Y|Z=z,X)|Z=z,X} \\
	&=E_v\bc{ \bb{Y - E_v\ba{Y|Z=z,X}}S_v(Y|Z=z,X)|Z=z,X}
	\\ \\
	&\nabla_{v}E_v\ba{Y|A=1,Z=z,X} \\
	&= \frac{\partial}{\partial v} \int Yf_v(Y|A=1,Z=z,X)dY \\
	&= \int Y\frac{f'_v(Y|A=1,Z=z,X)}{f_v(Y|A=1,Z=z,X)} f_v(Y|A=1,Z=z,X) dy\\ 
	&=E_v\bb{YS_v(Y|A=1,Z=z,X)|A=1,Z=z,X} \\
	&=E_v\bc{ \bb{Y - E_v\ba{Y|A=1,Z=z,X}}S_v(Y|A=1,Z=z,X)|A=1,Z=z,X}
\end{align*}
The last equality holds by T2.

$\nabla_{v}\delta_{Yv}(X)$ has the following representation:
\begin{align*}
	& \nabla_{v} \delta_{Yv}(X)\\
	&= \nabla_{v}E_v\ba{Y|Z=1,X} - \nabla_{v}E_v\ba{Y|Z=0,X} - \bc{\nabla_{v}E_v\ba{Y|A=1,Z=1,X} - \nabla_{v}E_v\ba{Y|A=1,Z=0,X}}\\
	&= 
	E_v \bc{ \bb{Y - E_v\ba{Y|Z=1,X}} S_v(Y,A|Z=1,X) |Z=1,X}
	 - 
	E_v \bc{ \bb{Y - E_v\ba{Y|Z=0,X}} S_v(Y,A|Z=0,X) |Z=0,X }   
    \\
    &\quad - E_v\bc{ \bb{Y - E_v\ba{Y|A=1,Z=z,X}}S_v(Y|A=1,Z=z,X)|A=1,Z=z,X} \\
    &\quad + E_v\bc{ \bb{Y - E_v\ba{Y|A=1,Z=z,X}}S_v(Y|A=1,Z=z,X)|A=1,Z=z,X} \\
	&= 
	E_v \bc{\frac{2Z-1}{f_v(Z|X)}\bb{Y - E_v\ba{Y|Z,X}}S_v(Y,A|Z,X) \Big|X } \\
	&\quad -
	E_v \bc{\frac{2Z-1}{f_v(Z|X)} \frac{A}{Pr_v(A=1|Z,X)}\bb{Y - E_v\ba{Y|A=1,Z,X}}S_v(Y,A|Z,X) \Big|X } 
    \\
	&= 
	E_v \bc{\frac{2Z-1}{f_v(Z|X)}
	\bc{
	\bb{Y - E_v\ba{Y|Z,X}} - \frac{A}{Pr_v(A=1|Z,X)}\bb{Y - E_v\ba{Y|A=1,Z,X}}
	}
	S_v(Y,A,Z|X) \Big|X }.
    \numberthis \label{eq-deltaY}
\end{align*}
The last equality holds by T1. 

Based on analogous algebra, one can show that:
\begin{align}
	&\nabla_{v} \delta_{Av}(X) = E_v\bc{\frac{2Z-1}{f_v(Z|X)} \{A - Pr_v(A=1|Z,X)\}S_v(Y,A,Z|X) \Big|X }
    \label{eq-deltaA}
\end{align}

Combining \eqref{eq-delta}-\eqref{eq-deltaA}, we have
\begin{align*}
	&\nabla_{v}\delta^*_v(X) \\
	&= \frac{1}{\delta_{Av}(X)^2}
	\bc{
	\bb{\nabla_{v} \delta_{Yv}(X)} \delta_{Av}(X) - \delta_{Yv}(X) \bb{\nabla_{v} \delta_{Av}(X)}	
	}\\
	&= 
	\frac{1}{\delta_{Av}(X)^2}
	\bc{ 
    \begin{array}{l} 
    E_v\bc{\frac{2Z-1}{f_v(Z|X)} \bc{
	\bb{Y - E_v\ba{Y|Z,X}} - \frac{A}{Pr_v(A=1|Z,X)}\bb{Y - E_v\ba{Y|A=1,Z,X}}
	}
	S_v(Y,A,Z|X) \Big|X }
	\delta_{Av}(X)   \\ 
	  \hspace*{1cm} -  \delta_{Yv}(X) E_v \bc{\frac{2Z-1}{f(Z|X)} \bb{A - Pr_v(A=1|Z,X)}S_v(Y,A,Z|X)\} \Big|X} 
    \end{array}}\\
	&=E_v\bc{ 
    \left.
    \begin{array}{l}
    \frac{1}{\delta_{Av}(X)}\frac{2Z-1}{f_v(Z|X)}
    \bc{
	\bb{Y - E_v\ba{Y|Z,X}} - \frac{A}{Pr_v(A=1|Z,X)}\bb{Y - E_v\ba{Y|A=1,Z,X}}
	}
    S_v(Y,A,Z|X) 
    \\
    \quad   - \frac{1}{\delta_{Av}(X)}\frac{2Z-1}{f(Z|X)}\bb{A - Pr_v(A=1|Z,X)}S_v(Y,A,Z|X) \delta^*_v(X)
    \end{array}
    \right| X 
    } \\
	&=E_v\bc{
	 \frac{2Z-1}{f_v(Z|X)} \frac{1}{\delta_{Av}(X)}
	\bc{
	\begin{aligned}
	&Y - E_v\ba{Y|Z,X} - \bb{A - Pr_v(A=1|Z,X)} \delta^*_v(X)\\
	&-\frac{A}{Pr_v(A=1|Z,X)}\bb{Y - E_v\ba{Y|A=1,Z,X}}
	\end{aligned}
	} S_v(Y,A,Z|X) \Big|X 
	}.
\end{align*}

Based on this result, Part 2 is expressed as follows:
\begin{align*}
	&E_v \Big[ \bb{\nabla_{v}\delta^*_v(X)} \frac{A}{Pr_v(A=1)} \Big] \\
	&=\int \bb{\nabla_{v}\delta^*_v(X)} f_v(X|A=1)dX  \\
	&=\int  \frac{ Pr_v(A=1|X)}{Pr_v(A=1)}  \bb{\nabla_{v}\delta^*_v(X)} f_v(X) dX \\
	&=\int \int \frac{ Pr_v(A=1|X)}{Pr_v(A=1)}\frac{2Z-1}{f_v(Z|X)} \frac{1}{\delta_{Av}(X)}
	\bc{
	\begin{aligned}
	&Y - E_v\ba{Y|Z,X} - \bb{A - Pr_v(A=1|Z,X)} \delta^*_v(X)\\
	&-\frac{A}{Pr_v(A=1|Z,X)}\bb{Y - E_v\ba{Y|A=1,Z,X}}
	\end{aligned}
	}  \\ 
	&\qq \qq  \times S_v(Y,A,Z|X) f_v(Y,A,Z|X) d(Y,A,Z) \ f_v(X) dX\\
	&=\int \frac{Pr_v(A=1|X)}{Pr_v(A=1)} 
	 \frac{2Z-1}{f_v(Z|X)} \frac{1}{\delta_{Av}(X)}	\bc{
	\begin{aligned}
	&Y - E_v\ba{Y|Z,X} - \bb{A - Pr_v(A=1|Z,X)} \delta^*_v(X)\\
	&-\frac{A}{Pr_v(A=1|Z,X)}\bb{Y - E_v\ba{Y|A=1,Z,X}}
	\end{aligned}
	} S_v(O) f_v(O) dO \\ 		 
	&=E_v\bc{
	 \frac{Pr_v(A=1|X)}{Pr_v(A=1)}  \frac{2Z-1}{f_v(Z|X)} \frac{1}{\delta_{Av}(X)}
	\bc{
	\begin{aligned}
	&Y - E_v\ba{Y|Z,X} - \bb{A - Pr_v(A=1|Z,X)} \delta^*_v(X)\\
	&-\frac{A}{Pr_v(A=1|Z,X)}\bb{Y - E_v\ba{Y|A=1,Z,X}}
	\end{aligned}
	}
	S_v(O)}
\end{align*}
The third-to-last equality holds by T2.

Define $\theta_v(O)$ as
$$\theta_v(O) :=  \frac{2Z-1}{f_v(Z|X)} \frac{1}{\delta_{Av}(X)}
	\bc{Y - E_v\ba{Y|Z,X} - \bb{A - Pr_v(A=1|Z,X)} \delta^*_v(X)-\frac{A}{Pr_v(A=1|Z,X)}\bb{Y - E_v\ba{Y|A=1,Z,X}}}
	 	$$
Combining the derived expressions for Part 1, Part 2 and Part 3, we establish the following result:
\begin{align*}
	&\frac{\partial}{\partial v} \delta^*_v \\
	&= \text{Part 1 + Part 2 - Part 3} \\
	&= E_v\bc{\delta^*_v(X) \frac{A}{Pr_v(A=1)} S_v(O) +\frac{Pr_v(A=1|X)}{Pr_v(A=1)}  \theta_v(O)  S_v(O) - \delta^*_v \frac{A}{Pr_v(A=1)} S_v(O) } \\
	&= E_v\bc{ \bc{ \frac{A}{Pr_v(A=1)} \ba{\delta^*_v(X) - \delta^*_v} + \theta_v(O) } S_v(O)}
\end{align*}
This result implies that $\delta^*$ is pathwise differentiable \cite{newey1990semiparametric} with an influence function $EIF(O;\delta^*)$ defined as follows:
\begin{align*}
	&EIF(O;\delta^*) = \frac{A}{Pr(A=1)} \bb{\delta^*(X) - \delta^*} + \frac{\rho(X)}{Pr(A=1)} \theta(O) , \\
	&\theta(O) := \frac{2Z-1}{f(Z|X)} \frac{1}{\delta_A(X)}
	\bc{ Y - E\ba{Y|Z,X} - \bb{A - Pr(A=1|Z,X)} \delta^*(X) -\frac{A}{Pr(A=1|Z,X)}\bb{Y - E\ba{Y|A=1,Z,X}}}.
\end{align*}
Of note, $EIF(O;\delta^*)$ derived above satisfies the following equation:
\begin{equation*}
 \frac{\partial}{\partial v} \delta^*_v \bigg|_{v=0}
 = E\bb{EIF(O;\delta^*) S(O)}.
\end{equation*}
In other words, $\delta^*$ is pathwise differentiable 
\cite{newey1990semiparametric}, with an influence function $EIF(O;\delta^*)$. Since the model $\mathcal{M}$ is nonparametric, $EIF(O;\delta^*)$ is the unique influence function satisfying \eqref{eq-EIF}. Consequently, $EIF(O;\delta^*)$ is the EIF for $\delta^*$. This completes the proof.

Of note, $\theta(O)$ has the alternative representation:
\begin{align*}
	&\theta(O) := \frac{2Z-1}{f(Z|X)} \frac{1}{\delta_A(X)}
	\bc{ Y - E\ba{Y|Z,X} - \bb{A - Pr(A=1|Z,X)} \delta^*(X) -\frac{A}{Pr(A=1|Z,X)}\bb{Y - E\ba{Y|A=1,Z,X}}}\\
	&=\frac{2Z-1}{f(Z|X)} \frac{1}{\delta_A(X)}
	\bc{
	\begin{aligned}
		&Y - Z\bb{E\ba{Y|Z=1,X}-E\ba{Y|Z=0,X}} - E\ba{Y|Z=0,X} \\
		& - \bb{A - Z\bb{Pr(A=1|Z=1,X) - Pr(A=1|Z=0,X)} - Pr(A=1|Z=0,X)} \delta^*(X) \\
		&-\frac{A}{Pr(A=1|Z,X)}\bb{Y - Z\bb{E\ba{Y|A=1,Z=1,X}-E\ba{Y|A=1,Z=0,X}} -  E\ba{Y|A=1,Z=0,X}}
	\end{aligned}
	}\\	
	&=\frac{2Z-1}{f(Z|X)} \frac{1}{\delta_A(X)}
	\bc{
	\begin{aligned}
		&Y - E\ba{Y|Z=0,X} \\
		& - \bb{A - Pr(A=1|Z=0,X)} \delta^*(X) \\
		& - Z\bb{E\ba{Y|Z=1,X}-E\ba{Y|Z=0,X}}  +  Z\bb{Pr(A=1|Z=1,X) - Pr(A=1|Z=0,X)}\delta^*(X) \\
		&-\frac{A}{Pr(A=1|Z,X)}\bb{Y - Z\phi(X) -  E\ba{Y|A=1,Z=0,X}}
	\end{aligned}
	}\\		
	&=\frac{2Z-1}{f(Z|X)} \frac{1}{\delta_A(X)}
	\bc{
	\begin{aligned}
		&Y - A\delta^*(X) - E\ba{Y|Z=0,X} + Pr(A=1|Z=0,X)\delta^*(X)\\
		&-\frac{A}{Pr(A=1|Z,X)}\bb{Y - Z\phi(X) -  E\ba{Y|A=1,Z=0,X}}
	\end{aligned}
	}\\		
	&=\frac{2Z-1}{f(Z|X)} \frac{1}{\delta_A(X)}
	\bc{
	\begin{aligned}
		&Y - A\delta^*(X) -  Z\phi(X)  - \bb{E\ba{Y|Z=0,X} - Pr(A=1|Z=0,X)\delta^*(X)}\\
		&-\frac{A}{Pr(A=1|Z,X)}\bb{Y - Z\phi(X) -  E\ba{Y|A=1,Z=0,X}}
	\end{aligned}
	}\\			
	&=\frac{2Z-1}{f(Z|X)} \frac{1}{\delta_A(X)}
	\bc{
	\begin{aligned}
		&Y - A\delta^*(X) -  Z\phi(X) -E\bb{Y-A\delta^*(X)-Z\phi(X)|X} \\
		&-\frac{A}{Pr(A=1|Z,X)}\bb{Y - Z\phi(X) -  E\ba{Y|A=1,Z=0,X}}
	\end{aligned}
	}
\end{align*}

The last equation can be seen from
\begin{align*}
	&E\bb{Y-A\delta^*(X)-Z\phi(X)|X} \\
	&=E\bc{
	 \bc{E\bb{Y-A\delta^*(X)-\phi(X)|Z=1,X}-E\bb{Y-A\delta^*(X)|Z=0,X}} Z
	 | X} + E\bb{Y-A\delta^*(X)|Z=0,X}\\
	&=
	E
	\bc{
	\left.
	\begin{aligned}
		&\bb{E\ba{Y|Z=1,X} - E\ba{Y|Z=0,X}}Z - \phi(X)Z
		\\
		&-\bb{Pr(A=1|Z=1,X)-Pr(A=1|Z=0,X)}\delta^*(X)Z
		\\
		&+ E\bb{Y|Z=0,X} - Pr(A=1|Z=0,X) \delta^*(X)
	\end{aligned}
	\right|
	X
	}
	\\
	&=
	E
	\bc{
	\left.
	\begin{aligned}
		&\bb{E\ba{Y|Z=1,X} - E\ba{Y|Z=0,X}}Z - \phi(X)Z
		\\
		&-\bb{E\ba{Y|Z=1,X} - E\ba{Y|Z=0,X} - \phi(X)}Z
		\\
		&+ E\bb{Y|Z=0,X} - Pr(A=1|Z=0,X) \delta^*(X)
	\end{aligned}
	\right|
	X
	}
	\\
	&=E\ba{Y|Z=0,X} - Pr(A=1|Z=0,X)\delta^*(X) 
\end{align*}

Use the notation in the maintext, one has
\begin{align*}
	\theta(O) = \frac{2Z-1}{f(Z|X)} \frac{1}{p_1(X)-p_0(X)}\bc{Y - A\delta^*(X) -  Z\phi(X) -w(X) - \frac{A}{p_Z(X)}\bb{Y - Z\phi(X) -  e_{10}(X)}};
\end{align*}
or equivalently,
\begin{align*}
	\theta(O) = \frac{2Z-1}{f(Z|X)} \frac{1}{p_1(X)-p_0(X)}\bc{Y - e_Z(X) -  \bb{A-p_Z(X)}\delta^*(X) - \frac{A}{p_Z(X)}\bb{Y - Z\phi(X) -  e_{10}(X)}},
\end{align*}

\textbf{\textit{Multiple Robustness Property}}

We verify that the EIF is an unbiased moment equation under each of the models $\mc{M}_1$, $\mc{M}_2$ and $\mc{M}_3$. Again, we let the superscript $\sim$ indicate a misspecified model.
\begin{align*}
    &\mc{M}_1: \textit{models for }  p_z(X) \textit{ and }  \pi_z(X) \textit{ are evaluated at their true value, for } z\in\{0,1\} \\
	& E\bb{EIF(O;\delta^*,\tilde{e}_{1z}(X),p_z(X),\pi_z(X), \tilde{w}(X), \tilde{\delta}^*(X))} \\
	&=
	E\bc{
	\frac{A}{Pr(A=1)}\bb{\tilde{\delta}^*(X) - \delta^*}
	} \\
	&\quad +
	E\bc{
	\frac{\rho(X)}{Pr(A=1)}
	\frac{2Z-1}{\pi_Z(X)} \frac{1}{p_1(X)-p_0(X)}
	\bc{Y - \tilde{e}_Z(X) - \bb{A-p_Z(X)}\tilde{\delta}^*(X) - \frac{A}{p_Z(X)}\bb{Y - Z\tilde{\phi}(X) -  \tilde{e}_{10}(X)}
	}
	}
	\\
	&=
	\ba{\tilde{\delta}^* - \delta^*}\\
	&\quad  +
	E\bc{
		\frac{\rho(X)}{Pr(A=1)}
		\frac{1}{p_1(X)-p_0(X)}
		\bc{
		e_1(X) - e_0(X) 
		- \bb{\tilde{e}_1(X)-\tilde{e}_0(X)} -
		\bb{e_{11}(X)-e_{10}(X) - \tilde{e}_{11}(X) -  \tilde{e}_{10}(X)}
		}
	}
	\\
	&=
	\ba{\tilde{\delta}^* - \delta^*}\\
	&\quad  +
	E\bc{
		\frac{\rho(X)}{Pr(A=1)}
		\frac{e_1(X) - e_0(X) - \bb{e_{11}(X)-e_{10}(X)}}{p_1(X)-p_0(X)}
	}	
	-
	E\bc{
		\frac{\rho(X)}{Pr(A=1)}
		\frac{\tilde{e}_1(X)-\tilde{e}_0(X)  - \bb{\tilde{e}_{11}(X) -  \tilde{e}_{10}(X)}}{p_1(X)-p_0(X)}
		}
	\\
	&=\ba{\tilde{\delta}^* - \delta^*} + \ba{\delta^*-\tilde{\delta}^*} \\
	&=0
\end{align*}
\begin{align*}
    &\mc{M}_2: \textit{models for } \delta^*(X), e_{1z}(X) \textit{ and } w(X)\textit{ are evaluated at their true value, for } z\in\{0,1\};\\
	& E\bb{EIF(O;\delta^*,e_{1z}(X),\tilde{p}_z(X),\tilde{\pi}_z(X), w(X), \delta^*(X))} \\
	&=
	E\bc{
	\frac{A}{Pr(A=1)}\bb{\delta^*(X) - \delta^*}
	} \\
	&\quad +
	E\bc{
	\frac{\tilde{\rho}(X)}{Pr(A=1)}
	\frac{2Z-1}{\tilde{\pi}_Z(X)} \frac{1}{\tilde{p}_1(X)-\tilde{p}_0(X)}
	\bc{Y - A\delta^*(X) - Z\phi(X) - w(X) - \frac{A}{\tilde{p}_Z(X)}\bb{Y - Z\phi(X) -  e_{10}(X)}
	}
	}
	\\
	&= 0  +
	E\bc{
	\frac{\tilde{\rho}(X)}{Pr(A=1)}
	\frac{2Z-1}{\tilde{\pi}_Z(X)} \frac{1}{\tilde{p}_1(X)-\tilde{p}_0(X)}
	\bc{0 - \frac{A}{\tilde{p}_Z(X)}\cdot 0}
	}
	\\
	&=0
\end{align*}
\begin{align*}
    &\mc{M}_3: \textit{models for } \delta^*(X), e_{1z}(X) \textit{ and }  \pi_z(X) \textit{ are evaluated at their true value, for } z\in\{0,1\} \\
	& E\bb{EIF(O;\delta^*,e_{1z}(X),\tilde{p}_z(X),\pi_z(X), \tilde{w}(X), \delta^*(X))} \\
	&=
	E\bc{
	\frac{A}{Pr(A=1)}\bb{\delta^*(X) - \delta^*}
	} \\
	&\quad +
	E\bc{
	\frac{\tilde{\rho}(X)}{Pr(A=1)}
	\frac{2Z-1}{\pi_Z(X)} \frac{1}{\tilde{p}_1(X)-\tilde{p}_0(X)}
	\bc{Y - \tilde{e}_Z(X) - \bb{A-\tilde{p}_Z(X)}\delta^*(X) - \frac{A}{\tilde{p}_Z(X)}\bb{Y - Z\phi(X) -  e_{10}(X)}
	}
	}
	\\
	&=
	E\bc{
		\frac{\tilde{\rho}(X)}{Pr(A=1)}
		\frac{1}{\tilde{p}_1(X)-\tilde{p}_0(X)}
		\bc{
		e_1(X) - e_0(X) 
		- \bb{\tilde{e}_1(X)-\tilde{e}_0(X)} -
		\bb{p_1(X)-p_0(X)}\delta^*(X) + \bb{\tilde{p}_1(X)-\tilde{p}_0(X)}\delta^*(X) 
		}
	}
	\\
	&=
	E\bc{
		\frac{\tilde{\rho}(X)}{Pr(A=1)}
		\frac{1}{\tilde{p}_1(X)-\tilde{p}_0(X)}
		\bc{
		\phi(X)
		- \bb{\tilde{e}_1(X)-\tilde{e}_0(X)} 
		+ \bb{\tilde{p}_1(X)-\tilde{p}_0(X)}\delta^*(X) 
		}
	}	
	\\
	&=
	E\bc{
		\frac{\tilde{\rho}(X)}{Pr(A=1)}
		\bc{
		\phi(X)
		- \bc{\bb{\tilde{p}_1(X)-\tilde{p}_0(X)}\delta^*(X) +\phi(X)}
		+ \bb{\tilde{p}_1(X)-\tilde{p}_0(X)}\delta^*(X) 
		}
	}
	\\	
	&=0
\end{align*}

\section{Proof of Asymptotic Normality}
\subsection[Proof of \Cref{th:asym}                         ]{Proof of \Cref{th:asym} -- Asymptotic Normality of the Proposed Estimator}
In this section, we prove \Cref{th:asym} in the main text. For notational convenience, we use the shorthand \(f=f(X)\) or \(f=f(O)\), and apply this convention to any function of \(X\), including \(e_z=e_z(X)\), \(e_{1z}=e_{1z}(X)\), \(p_z=p_z(X)\), \(\pi_z=\pi_z(X)\), \(\rho=\rho(X)\), \(\delta^*=\delta^*(X)\), \(w=w(X)\) and \(\theta=\theta(X)\). To avoid confusion between the conditional quantity \(\delta^*(X)\) and its marginal counterpart, we re-denote the marginal estimand by
$$
\delta_m^* := E\bb{\frac{A}{Pr(A=1)}\,\delta^*(X)}.
$$
Throughout the remainder of this section, \(\delta_m^*\) refers to the marginal quantity, whereas \(\delta^*\) always denotes the conditional one.
\\ \\
If one can show that $\hat{\delta}^{*(k)}$ has the asymptotic representation 
\begin{align*}
	\sqrt{|\mc{I}_k|} \ba{\hat{\delta}^{*(k)} - \delta^*_m} = \frac{1}{\sqrt{|\mc{I}_k|}} \sum_{i \in \mc{I}_k} EIF\ba{O_i;\delta^*_m} + o_P(1), 
	\numberthis \label{e-single}
\end{align*}
and one further establishes $$\hat{\delta}^{*EIF} = K^{-1} \sum_{k=1}^K \hat{\delta}^{*(k)}_m$$ has asymptotic representation
\begin{align*}
	\sqrt{N} \ba{\hat{\delta}^{*EIF} - \delta^*_m} = \frac{1}{\sqrt{N}} \sum_{i=1}^N EIF\ba{O_i;\delta^*_m} + o_P(1). 
	\numberthis \label{e-merged}
\end{align*}
Then, the asymptotic normality result holds from a standard central limit theorem. Hence, we begin by proving that \eqref{e-single} holds under assumptions \ref{as:iv1} -- \ref{as:rates}.
\\ \\
Recall 
\begin{align*}
	&\hat{\delta}^{*(k)} = \bb{\mb{P}(A)}^{-1}
	\mb{P}_{\mc{I}_k}
	\bc{
	 A\hat{\delta}^{*(-k)}+ \hat{\theta}^{(-k)}
     }, \\
	& 
	\hat{\theta}^{(-k)} = 
    \frac{\hat{\rho}^{(-k)}}{\hat{p}^{{(-k)}}_1-\hat{p}^{{(-k)}}_0} \frac{2Z-1}{ \hat{\pi}^{(-k)}_Z}  
    \bc{
    Y-A\hat{\delta}^{*(-k)}-Z\hat{\phi}^{(-k)} - \hat{w}^{(-k)} - \frac{A}{\hat{p}^{{(-k)}}_Z} \bb{Y-Z\hat{\phi}^{(-k)} + \hat{e}^{(-k)}_{10}}
    }	
	, \\
	&\hat{\rho}^{(-k)}  = \hat{p}_1^{(-k)} \hat{\pi}_1^{(-k)} + \hat{p}_0^{(-k)} \hat{\pi}_0^{(-k)}.
\end{align*}
Let $M_i=  A_i\delta_i^* + \theta_i$ and $\widehat{M}_i^{(-k)}=  A_i\hat{\delta}^{*(-k)}_i + \hat{\theta}^{(-k)}_i$. The left-hand side of \eqref{e-single} is given by
\begin{align*}
	&\sqrt{|\mc{I}_k|} \ba{\hat{\delta}^{*(k)} - \delta^*_m}  \\
	&=\frac{1}{\sqrt{|\mc{I}_k|}} 
	\sum_{i \in \mc{I}_k}
	\bb{\frac{\widehat{M}_i^{(-k)}}{\mb{P}(A)} - \frac{A_i\delta^*_m}{\mb{P}_{\mc{I}_k}(A)}} \\
	&=\frac{1}{\sqrt{|\mc{I}_k|}} 
	\sum_{i \in \mc{I}_k}
	\bb{\frac{Pr(A=1)}{\mb{P}(A)}\frac{\widehat{M}_i^{(-k)}}{Pr(A=1)} - \frac{Pr(A=1)}{\mb{P}_{\mc{I}_k}(A)}\frac{A_i\delta^*_m}{Pr(A=1)}} 
	\\	
	&\qq \textit{(use the decomposition that $ab-cd = \frac{1}{2}(a-c)(b+d) + \frac{1}{2}(a+c)(b-d)$)} \\
	&=\frac{1}{2}\frac{1}{\sqrt{|\mc{I}_k|}} 
	\underbrace{	
	\bb{
	\frac{Pr(A=1)}{\mb{P}(A)} - \frac{Pr(A=1)}{\mb{P}_{\mc{I}_k}(A)}
	}
	}_{o_p(1)} 
	\underbrace{	
	\sum_{i \in \mc{I}_k} \frac{\widehat{M}_i^{(-k)} + A_i\delta^*_m}{Pr(A=1)}
	}_{O_p(1)} 	
	+
	\frac{1}{2}\frac{1}{\sqrt{|\mc{I}_k|}}
	\underbrace{		
	\bb{
	\frac{Pr(A=1)}{\mb{P}(A)} + \frac{Pr(A=1)}{\mb{P}_{\mc{I}_k}(A)}
	}
	}_{2+o_p(1)} 	
	\underbrace{		
	\sum_{i \in \mc{I}_k} \frac{\widehat{M}_i^{(-k)} - A_i\delta^*_m}{Pr(A=1)}	
	}_{AN: O_p(1)} 		
	\\
	&=\frac{1}{\sqrt{|\mc{I}_k|}} \sum_{i \in \mc{I}_k}\bb{\frac{\widehat{M}_i^{(-k)}-A_i\delta^*_m}{Pr(A=1)}} + o_p(1)
\end{align*}
The fourth line holds because $\mb{P}_{\mc{I}_k}(A) = Pr(A=1)+o_P(1)$ and $\mb{P}(A) = Pr(A=1)+o_P(1)$ by the law of large number and $\widehat{M}_i^{(-k)}$ is bounded.
\\ \\
Let \( \mb{G}_{\mc{I}_k}(V) = |\mc{I}_k|^{-1/2} \sum_{i \in \mc{I}_k} \left\{ V_i - E(V_i) \right\} \) be the empirical process of \( V_i \) centered by \( E(V_i) \).  Similarly, let \( \mb{G}^{(-k)}_{\mc{I}_k}\ba{\widehat{V}^{(-k)}} = |\mc{I}_k|^{-1/2} \sum_{i \in \mc{I}_k} \bb{ \widehat{V}^{(-k)}_i - E^{(-k)}\ba{\widehat{V}^{(-k)}} } \) be the empirical process of \( \widehat{V}^{(-k)} \) centered by \( E^{(-k)}\ba{\widehat{V}^{(-k)}} \),  where \( E^{(-k)}(\cdot) \) is the expectation after considering random functions obtained from \( \mc{I}_k^c \) as fixed functions.  
\\ \\
The empirical process of \( \widehat{M}_i^{(-k)} - A_i\delta^*_m \) is
\begin{align}
	&
	|\mc{I}_k|^{-1/2} 
	\sum_{i \in \mc{I}_k}
	\bb{
	\widehat{M}_i^{(-k)}-A_i\delta^*_m
	}
	\notag
	\\
	&=
	\mb{G}_{\mc{I}_k}\ba{M - A\delta^*_m}
	\label{b-term1}
	\\
	&\quad + 
	 |\mc{I}_k|^{1/2} E^{(-k)} \ba{\widehat{M}_i^{(-k)} - M}
	 \label{b-term2}	
	\\
	&\quad + \mb{G}^{(-k)}_{\mc{I}_k} \ba{\widehat{M}^{(-k)} - M} \label{b-term3}		
\end{align}
From the two subsections below, we show that (\ref{b-term2}) and (\ref{b-term3}) are $o_P(1)$, indicating that (\ref{b-term1}) is asymptotically normal, and consequently, $(AN)$ is $O_p(1)$. Moreover, we establish (\ref{e-single}), concluding the proof as follows:
\begin{align*}
	\sqrt{|\mc{I}_k|} \ba{\hat{\delta}^{*(k)}_m - \delta^*_m} = \frac{1}{\sqrt{|\mc{I}_k|}} \sum_{i \in \mc{I}_k}\bb{\frac{\widehat{M}_i^{(-k)}-A_i\delta^*_m}{Pr(A=1)}} + o_p(1) 
	= \frac{1}{\sqrt{|\mc{I}_k|}} \sum_{i \in \mc{I}_k} EIF(O_i;\delta^*_m) + o_p(1).
\end{align*}

\textbf{\textit{Asymptotic Property of (\ref{b-term2})}}
\\ \\
The term (\ref{b-term2})
\begin{align*}
	&|\mc{I}_k|^{1/2} E^{(-k)} \ba{\widehat{M}^{(-k)} - M} \\
	&= 
	|\mc{I}_k|^{1/2} E^{(-k)} \bb{
	A\hat{\delta}^{*(-k)} + \hat{\theta}^{(-k)} -  A\delta^* - \theta
	}
	\\
	&= 
	|\mc{I}_k|^{1/2} E^{(-k)} \bb{
	A\ba{\hat{\delta}^{*(-k)} - \delta^*} + \hat{\theta}^{(-k)} -  \theta
	}	
	\\
	&= 
	|\mc{I}_k|^{1/2} E^{(-k)} \bb{
	A\ba{\hat{\delta}^{*(-k)} - \delta^*} + \hat{\theta}^{(-k)} -  \theta
	}.
\end{align*}
We note $E^{(-k)}\ba{\theta} =0$ (mean 0), $A$ in $E^{(-k)}\bb{A\ba{\hat{\delta}^*-\delta^*}}$ contributes $\rho=Pr(A=1|X)$. Thus, one has the decomposition that
\begin{align*}
	&E^{(-k)}\ba{\hat{\theta}^{(-k)}} \\
	&=E^{(-k)}\bb{
	\frac{\hat{\rho}^{(-k)}}{\hat{p}^{{(-k)}}_1-\hat{p}^{{(-k)}}_0} \frac{2Z-1}{ \hat{\pi}^{(-k)}_Z}  
    \bc{
    Y-A\hat{\delta}^{*(-k)}-Z\hat{\phi}^{(-k)} - \hat{w}^{(-k)} - \frac{A}{\hat{p}^{{(-k)}}_Z} \bb{Y-Z\hat{\phi}^{(-k)} + \hat{e}^{(-k)}_{10}}
    }	
	}
	\\
	&\textit{apply the conditional expectation that $Y = (e_1 - e_0)Z+e_0$ and $A = (p_1-p_0)Z+p_0$}
	\\
	&=E^{(-k)}\bb{
	\frac{\hat{\rho}^{(-k)}}{\hat{p}^{{(-k)}}_1-\hat{p}^{{(-k)}}_0} \frac{2Z-1}{ \hat{\pi}^{(-k)}_Z}  
    \bc{
    \begin{aligned}
    	&\ba{e_1-e_0}Z+e_0 - \bb{(p_1-p_0)Z+p_0}\hat{\delta}^{*(-k)} - Z\hat{\phi}^{(-k)} - \hat{w}^{(-k)} \\
    	&- \frac{A}{\hat{p}^{{(-k)}}_Z} \bb{Y-Z\hat{\phi}^{(-k)} + \hat{e}^{(-k)}_{10}}
    \end{aligned}
    }	
	}	\\
	&=E^{(-k)}\bb{
	\frac{\hat{\rho}^{(-k)}}{\hat{p}^{{(-k)}}_1-\hat{p}^{{(-k)}}_0} \frac{2Z-1}{ \hat{\pi}^{(-k)}_Z}  
    \bc{
    \begin{aligned}
    	&Z \delta^*(p_1-p_0) + Z\phi  + e_0 - Z(p_1-p_0)\hat{\delta}^{*(-k)} - p_0\hat{\delta}^{*(-k)}\\
    	&- Z\hat{\phi}^{(-k)} - \hat{w}^{(-k)} - \frac{A}{\hat{p}^{{(-k)}}_Z}
    	 \bb{Z(e_{11}-e_{10})+e_{10}-Z\hat{\phi}^{(-k)} + \hat{e}^{(-k)}_{10}}
    \end{aligned}
    }	
	}	\\
	&=E^{(-k)}\bb{
	\frac{\hat{\rho}^{(-k)}}{\hat{p}^{{(-k)}}_1-\hat{p}^{{(-k)}}_0} \frac{2Z-1}{ \hat{\pi}^{(-k)}_Z}  
    \bc{
    \begin{aligned}
    	&Z (p_1-p_0) \ba{\delta^*-\hat{\delta}^{*(-k)}} + Z\ba{\phi-\hat{\phi}^{(-k)}}
    	\\
    	&+p_0\ba{\delta^* -\hat{\delta}^{*(-k)}} + \bb{e_0 - p_0\delta^* - \hat{w}^{(-k)}} \\
    	&+\bb{Z\ba{\phi-\hat{\phi}^{(-k)}} - \bb{e_{10}-\hat{e}^{(-k)}_{10}}}
    \end{aligned}
    }	
	}	
	\\
	&=E^{(-k)}\bb{
	\frac{\hat{\rho}^{(-k)}}{\hat{p}^{{(-k)}}_1-\hat{p}^{{(-k)}}_0} \frac{2Z-1}{ \hat{\pi}^{(-k)}_Z}  
    \bc{
    \begin{aligned}
    	&Z (p_1-p_0) \ba{\delta^*-\hat{\delta}^{*(-k)}} + Z\ba{\phi-\hat{\phi}^{(-k)}}
    	\\
    	&+p_0\ba{\delta^* -\hat{\delta}^{*(-k)}} + \bb{w - \hat{w}^{(-k)}} \\
    	&+\bb{Z\ba{\phi-\hat{\phi}^{(-k)}} - \bb{e_{10}-\hat{e}^{(-k)}_{10}}}
    \end{aligned}
    }	
	}.		
\end{align*}
The last equation holds by 
\begin{align*}
	&E^{(-k)}\ba{w} = E^{(-k)}\ba{Y-A\delta^*-Z\phi} = E^{(-k)}\bb{ \ba{e_1-e_0}Z+e_0 - A\delta^* -Z\phi } \\
    &= E^{(-k)}\bb{\ba{e_1-e_0}Z+e_0 - \bb{(p_1-p_0)Z+p_0}\delta^* -Z\phi}  \\
	&= E^{(-k)}\bb{Z(p_1-p_0)\delta^* +Z\phi + e_0 - Z(p_1-p_0)\delta^*-p_0\delta^* -Z\phi} = E^{(-k)}\ba{e_0-p_0\delta^*}.
\end{align*}
Then, the term (\ref{b-term2}) becomes
\begin{align*}
	&|\mc{I}_k|^{1/2} E^{(-k)} \ba{\widehat{M}^{(-k)} - M} \\
	&= 
	|\mc{I}_k|^{1/2} E^{(-k)} \bb{
	A\ba{\hat{\delta}^{*(-k)} - \delta^*} + \hat{\theta}^{(-k)} -  \theta
	}
	\\	
	&= 
	|\mc{I}_k|^{1/2} E^{(-k)} \bb{
	\rho
	\ba{\hat{\delta}^{*(-k)} - \delta^*} + \hat{\theta}^{(-k)} 
	}\\
	&= 
	|\mc{I}_k|^{1/2} E^{(-k)} \bc{
	\rho
	\ba{\hat{\delta}^{*(-k)} - \delta^*} + 
	\frac{\hat{\rho}^{(-k)}}{\hat{p}^{{(-k)}}_1-\hat{p}^{{(-k)}}_0} \frac{2Z-1}{ \hat{\pi}^{(-k)}_Z}  
    \bc{
    Y-A\hat{\delta}^{*(-k)}-Z\hat{\phi}^{(-k)} - \hat{w}^{(-k)} - \frac{A}{\hat{p}^{{(-k)}}_Z} \bb{Y-Z\hat{\phi}^{(-k)} + \hat{e}^{(-k)}_{10}}
    }
	}\\
	&= 
	|\mc{I}_k|^{1/2} E^{(-k)} 
	\bc{
	\begin{aligned}
		&\rho \ba{\hat{\delta}^{*(-k)} - \delta^*} 
		\\ 
		&+
		\frac{\hat{\rho}^{(-k)}}{\hat{p}^{{(-k)}}_1-\hat{p}^{{(-k)}}_0} \frac{2Z-1}{ \hat{\pi}^{(-k)}_Z} 
		\bc{
    \begin{aligned}
    	&Z (p_1-p_0) \ba{\delta^*-\hat{\delta}^{*(-k)}} + Z\ba{\phi-\hat{\phi}^{(-k)}}
    	\\
    	&+p_0\ba{\delta^* -\hat{\delta}^{*(-k)}} + \ba{w - \hat{w}^{(-k)}} \\
    	&+ \frac{A}{\hat{p}^{{(-k)}}}\bb{Z\ba{\phi-\hat{\phi}^{(-k)}} - \bb{e_{10}-\hat{e}^{(-k)}_{10}}}
    \end{aligned}		
		}
	\end{aligned}
	}\\
	&= 
	|\mc{I}_k|^{1/2} E^{(-k)} 
	\bc{
	\begin{aligned}
		&\rho \ba{\hat{\delta}^{*(-k)} - \delta^*} +
		\\ 
		&+
		\hat{\rho}^{(-k)}
		\frac{p_1-p_0}{\hat{p}^{{(-k)}}_1-\hat{p}^{{(-k)}}_0} \frac{\pi_1}{\hat{\pi}^{(-k)}_1}\ba{\delta^*-\hat{\delta}^{*(-k)}} 
		\\
		&+
		\hat{\rho}^{(-k)}
		\frac{1}{\hat{p}^{{(-k)}}_1-\hat{p}^{{(-k)}}_0} \frac{\pi_1}{\hat{\pi}^{(-k)}_1}\ba{\phi-\hat{\phi}^{(-k)}}	
		\\
		&+
		\hat{\rho}^{(-k)}
		\frac{1}{\hat{p}^{{(-k)}}_1-\hat{p}^{{(-k)}}_0} \bb{\frac{\pi_1}{\hat{\pi}^{(-k)}_1}-\frac{\pi_0}{\hat{\pi}^{(-k)}_0}}\bb{
		\ba{w - \hat{w}^{(-k)}} + p_0\ba{\delta^* -\hat{\delta}^{*(-k)}}
		}
		\\
		&-
		\hat{\rho}^{(-k)}
		\frac{1}{\hat{p}^{{(-k)}}_1-\hat{p}^{{(-k)}}_0} \bb{\frac{\pi_1}{\hat{\pi}^{(-k)}_1}\frac{p_1}{\hat{p}^{{(-k)}}_1}}\ba{\phi-\hat{\phi}^{(-k)}}
		\\
		&-
		\hat{\rho}^{(-k)}
		\frac{1}{\hat{p}^{{(-k)}}_1-\hat{p}^{{(-k)}}_0} \bb{\frac{\pi_1}{\hat{\pi}^{(-k)}_1}\frac{p_1}{\hat{p}^{{(-k)}}_1}-\frac{\pi_0}{\hat{\pi}^{(-k)}_0}\frac{p_0}{\hat{p}^{{(-k)}}_0}}\bb{e_{10}-\hat{e}^{(-k)}_{10}}
	\end{aligned}
	}	
	.	
\end{align*}
We handle each term of the above expression:
\begin{align*}
	&\rho = p_1 \pi_1 + p_0 \pi_0 \\
	&|\mc{I}_k|^{1/2} E^{(-k)} \ba{\widehat{M}^{(-k)} - M} = |\mc{I}_k|^{1/2} E^{(-k)} \ba{\textrm{T1} + \textrm{T2} + \textrm{T3} + \textrm{T4} - \textrm{T5} - \textrm{T6}}
	\\
	& \text{T1} = (p_1\pi_1 + p_0\pi_0) \ba{\hat{\delta}^{*(-k)} - \delta^*}  \\
	& \text{T2} = 
	\ba{\hat{p}_1^{(-k)}\hat{\pi}_1^{(-k)} + \hat{p}_0^{(-k)}\hat{\pi}_0^{(-k)}}
	\frac{p_1-p_0}{\hat{p}^{{(-k)}}_1-\hat{p}^{{(-k)}}_0} \frac{\pi_1}{\hat{\pi}^{(-k)}_1}\ba{\delta^*-\hat{\delta}^{*(-k)}} 
	\\
	& \text{T3} = 
	\ba{\hat{p}_1^{(-k)}\hat{\pi}_1^{(-k)} + \hat{p}_0^{(-k)}\hat{\pi}_0^{(-k)}}
	\frac{1}{\hat{p}^{{(-k)}}_1-\hat{p}^{{(-k)}}_0} \frac{\pi_1}{\hat{\pi}^{(-k)}_1}\ba{\phi-\hat{\phi}^{(-k)}}	 
	\\ 
	&\text{T4} = 
	\ba{\hat{p}_1^{(-k)}\hat{\pi}_1^{(-k)} + \hat{p}_0^{(-k)}\hat{\pi}_0^{(-k)}}
	\frac{1}{\hat{p}^{{(-k)}}_1-\hat{p}^{{(-k)}}_0} \bb{\frac{\pi_1}{\hat{\pi}^{(-k)}_1}-\frac{\pi_0}{\hat{\pi}^{(-k)}_0}}\bb{\ba{w - \hat{w}^{(-k)}} + p_0\ba{\delta^* -\hat{\delta}^{*(-k)}}}
	\\ 
	&\text{T5} = 
	\ba{\hat{p}_1^{(-k)}\hat{\pi}_1^{(-k)} + \hat{p}_0^{(-k)}\hat{\pi}_0^{(-k)}}
	\frac{1}{\hat{p}^{{(-k)}}_1-\hat{p}^{{(-k)}}_0} \bb{\frac{\pi_1}{\hat{\pi}^{(-k)}_1}\frac{p_1}{\hat{p}^{{(-k)}}_1}}\ba{\phi-\hat{\phi}^{(-k)}}
	\\
	&\text{T6} = 
	\ba{\hat{p}_1^{(-k)}\hat{\pi}_1^{(-k)} + \hat{p}_0^{(-k)}\hat{\pi}_0^{(-k)}}
	\frac{1}{\hat{p}^{{(-k)}}_1-\hat{p}^{{(-k)}}_0} \bb{\frac{\pi_1}{\hat{\pi}^{(-k)}_1}\frac{p_1}{\hat{p}^{{(-k)}}_1}-\frac{\pi_0}{\hat{\pi}^{(-k)}_0}\frac{p_0}{\hat{p}^{{(-k)}}_0}}\bb{e_{10}-\hat{e}^{(-k)}_{10}}	.
\end{align*}
We focus on T1 first.
\begin{align*}
	&\text{T1} \\
	&= (p_1\pi_1 + p_0\pi_0) \ba{\hat{\delta}^{*(-k)} - \delta^*}  \\
	&= p_1 \ba{\pi_1-\hat{\pi}_1^{(-k)}}\ba{\hat{\delta}^{*(-k)} - \delta^*} + p_0 \ba{\pi_0-\hat{\pi}_0^{(-k)}}\ba{\hat{\delta}^{*(-k)} - \delta^*} \\
	&\quad 	+ p_1 \hat{\pi}_1^{(-k)}\ba{\hat{\delta}^{*(-k)} - \delta^*} + p_0\hat{\pi}_0^{(-k)}\ba{\hat{\delta}^{*(-k)} - \delta^*} \\
	&= p_1 \frac{\rho}{\rho}\ba{\pi_1-\hat{\pi}_1^{(-k)}}\ba{\hat{\delta}^{*(-k)} - \delta^*} 
    + p_0 \frac{\rho}{\rho}\ba{\pi_0-\hat{\pi}_0^{(-k)}}\ba{\hat{\delta}^{*(-k)} - \delta}
	\tag{T1-1} \\
	&\quad + \hat{\pi}_1^{(-k)} \frac{\rho}{\rho} \ba{p_1 -\hat{p}_1^{(-k)}} \ba{\hat{\delta}^{*(-k)} - \delta^*} + \frac{\rho}{\rho}\hat{\pi}_0 \ba{p_0-\hat{p}_0^{(-k)} } \ba{\hat{\delta}^{*(-k)} - \delta^*} 	\tag{T1-2} \\
	& \quad + \ba{\hat{p}_1^{(-k)}\hat{\pi}_1^{(-k)}  + \hat{p}_0^{(-k)}\hat{\pi}_0^{(-k)}}  \ba{\hat{\delta}^{*(-k)} - \delta}	\tag{T1-3} 
	\\ \\
	& E^{(-k)}\ba{\text{T1-1}} \lesssim \norm{\hat{\delta}^{*(-k)} - \delta^*} \cdot  \norm{\hat{\pi}_1^{(-k)}  - \pi_1} + \norm{\hat{\delta}^{*(-k)} - \delta^*}  \cdot  \norm{\hat{\pi}_0^{(-k)} - \pi_0} \\
	& E^{(-k)}\ba{\text{T1-2}} \lesssim \norm{\hat{\delta}^{*(-k)} - \delta^*} \cdot  \norm{\hat{p}_1^{(-k)} - p_1} + \norm{\hat{\delta}^{*(-k)} - \delta^*} \cdot  \norm{\hat{p}_0^{(-k)} -p_0}
\end{align*}
We consider T1-3 and T2 together.
\begin{align*}
	& \text{T1-3} +\text{T2} \\
	&= \ba{\hat{p}_1^{(-k)}\hat{\pi}_1^{(-k)}  + \hat{p}_0^{(-k)}\hat{\pi}_0^{(-k)}}  
	\ba{\hat{\delta}^{*(-k)} - \delta^*} 
	+
\ba{\hat{p}_1^{(-k)}\hat{\pi}_1^{(-k)} + \hat{p}_0^{(-k)}\hat{\pi}_0^{(-k)}}	
\frac{p_1-p_0}{\hat{p}^{{(-k)}}_1-\hat{p}^{{(-k)}}_0} \frac{\pi_1}{\hat{\pi}^{(-k)}_1}\ba{\delta^*-\hat{\delta}^{*(-k)}} 
	 \\
	&=
	\hat{\rho}^{(-k)} \frac{\hat{\pi}^{(-k)}_1}{\hat{\pi}^{(-k)}_1} \frac{\hat{p}^{{(-k)}}_1-\hat{p}^{{(-k)}}_0}{\hat{p}^{{(-k)}}_1-\hat{p}^{{(-k)}}_0}\ba{\hat{\delta}^{*(-k)} - \delta^*} 
	+
	\hat{\rho}^{(-k)} \frac{p_1-p_0}{\hat{p}^{{(-k)}}_1-\hat{p}^{{(-k)}}_0} \frac{\pi_1}{\hat{\pi}^{(-k)}_1}\ba{\delta^*-\hat{\delta}^{*(-k)}} 
	\\
	&=
	\hat{\rho}^{(-k)} \frac{\hat{\pi}^{(-k)}_1}{\hat{\pi}^{(-k)}_1} \frac{\hat{p}^{{(-k)}}_1-\hat{p}^{{(-k)}}_0}{\hat{p}^{{(-k)}}_1-\hat{p}^{{(-k)}}_0}\ba{\hat{\delta}^{*(-k)} - \delta^*} 
	+
	\hat{\rho}^{(-k)} \frac{p_1-p_0}{\hat{p}^{{(-k)}}_1-\hat{p}^{{(-k)}}_0} \frac{\pi_1}{\hat{\pi}^{(-k)}_1}\ba{\delta^*-\hat{\delta}^{*(-k)}} 	
	\\
	&=
	\hat{\rho}^{(-k)} \frac{1}{\hat{\pi}^{(-k)}_1} \frac{1}{\hat{p}^{{(-k)}}_1-\hat{p}^{{(-k)}}_0}\ba{\hat{\delta}^{*(-k)} - \delta^*} 		
	\underbrace{\bb{\hat{\pi}^{(-k)}_1\ba{\hat{p}^{{(-k)}}_1-\hat{p}^{{(-k)}}_0} - \pi_1\ba{p_1 - p_0}}}_{(*)}
	\\ 
	& \textit{apply the decomposition that $ab-cd = \frac{1}{2}(a-c)(b+d) + \frac{1}{2}(a+c)(b-d)$ to $(*)$}
	\\ 
	&E^{(-k)}\ba{\text{T1-3 + T2}} \lesssim
	\begin{aligned}
		&\norm{\hat{\delta}^{*(-k)}-\delta^*} \cdot \norm{\hat{\pi}^{(-k)}_1 - \pi_1} +
		\norm{\hat{\delta}^{*(-k)}-\delta^*} \cdot \norm{\hat{p}^{(-k)}_1 - p_1}
		\\
		&\qq \qq +\norm{\hat{\delta}^{*(-k)}-\delta^*} \cdot \norm{\hat{p}^{(-k)}_0 - p_0}
	\end{aligned}.
\end{align*}
Next, we handle T3$-$T5 toghether.
\begin{align*}
	&\text{T3$-$T5} \\ 
	&=\ba{\hat{p}_1^{(-k)}\hat{\pi}_1^{(-k)} + \hat{p}_0^{(-k)}\hat{\pi}_0^{(-k)}}
	\frac{1}{\hat{p}^{{(-k)}}_1-\hat{p}^{{(-k)}}_0} \frac{\pi_1}{\hat{\pi}^{(-k)}_1}\ba{\phi-\hat{\phi}^{(-k)}}	 	 \\
	&\qq - \ba{\hat{p}_1^{(-k)}\hat{\pi}_1^{(-k)} + \hat{p}_0^{(-k)}\hat{\pi}_0^{(-k)}}
	\frac{1}{\hat{p}^{{(-k)}}_1-\hat{p}^{{(-k)}}_0} \bb{\frac{\pi_1}{\hat{\pi}^{(-k)}_1}\frac{p_1}{\hat{p}^{{(-k)}}_1}}\ba{\phi-\hat{\phi}^{(-k)}}
	\\
	&=\hat{\rho}^{{(-k)}}
	\frac{1}{\hat{p}^{{(-k)}}_1-\hat{p}^{{(-k)}}_0} \frac{\pi_1}{\hat{\pi}^{(-k)}_1}
	\frac{\hat{p}^{{(-k)}}_1}{\hat{p}^{{(-k)}}_1}\ba{\phi-\hat{\phi}^{(-k)}}
	-
	\hat{\rho}^{{(-k)}}
	\frac{1}{\hat{p}^{{(-k)}}_1-\hat{p}^{{(-k)}}_0} \bb{\frac{\pi_1}{\hat{\pi}^{(-k)}_1}\frac{p_1}{\hat{p}^{{(-k)}}_1}}\ba{\phi-\hat{\phi}^{(-k)}}	
	\\
	&=\hat{\rho}^{{(-k)}}
	\frac{1}{\hat{p}^{{(-k)}}_1-\hat{p}^{{(-k)}}_0} \frac{\pi_1}{\hat{\pi}^{(-k)}_1}
	\frac{1}{\hat{p}^{{(-k)}}_1}\ba{\hat{p}^{{(-k)}}_1 - p_1}\ba{\phi-\hat{\phi}^{(-k)}}
	\\
	&E^{(-k)}\ba{\textrm{T3$-$T5}} \lesssim \norm{\hat{p}^{{(-k)}}_1 - p_1}\cdot \norm{\hat{\phi}^{(-k)} - \phi}.
\end{align*}
For T4, 
\begin{align*}
	&\textrm{T4} \\
	&= \ba{\hat{p}_1^{(-k)}\hat{\pi}_1^{(-k)} + \hat{p}_0^{(-k)}\hat{\pi}_0^{(-k)}}
	\frac{1}{\hat{p}^{{(-k)}}_1-\hat{p}^{{(-k)}}_0} \bb{\frac{\pi_1}{\hat{\pi}^{(-k)}_1}-\frac{\pi_0}{\hat{\pi}^{(-k)}_0}}\bb{\ba{w - \hat{w}^{(-k)}} + p_0\ba{\delta^* -\hat{\delta}^{*(-k)}}}
	\\
	&=\ba{\hat{p}_1^{(-k)}\hat{\pi}_1^{(-k)} + \hat{p}_0^{(-k)}\hat{\pi}_0^{(-k)}}
	\frac{1}{\hat{p}^{{(-k)}}_1-\hat{p}^{{(-k)}}_0} \frac{1}{\hat{\pi}^{(-k)}_1\hat{\pi}^{(-k)}_0}
	\underbrace{\bb{\hat{\pi}^{(-k)}_0\pi_1 - \hat{\pi}^{(-k)}_1 \pi_0}}_{(*)}
	\bb{\ba{w - \hat{w}^{(-k)}} + p_0\ba{\delta^* -\hat{\delta}^{*(-k)}}}
	\\ 
	& \textit{apply the decomposition that $ab-cd = \frac{1}{2}(a-c)(b+d) + \frac{1}{2}(a+c)(b-d)$ to $(*)$}
	\\ 
	&E^{(-k)}\ba{\textrm{T4}} \lesssim
	\begin{aligned}
		&\norm{\hat{\pi}^{(-k)}_0 - \pi_0} \cdot \norm{\hat{w}^{(-k)} - w} 
		+
		\norm{\hat{\pi}^{(-k)}_0 - \pi_0} \cdot \norm{\hat{\delta}^{*(-k)}-\delta^*} 
		\\
		&+\norm{\hat{\pi}^{(-k)}_1 - \pi_1} \cdot \norm{\hat{w}^{(-k)} - w} 
		+
		\norm{\hat{\pi}^{(-k)}_1 - \pi_1} \cdot \norm{\hat{\delta}^{*(-k)}-\delta^*} 
	\end{aligned}.
\end{align*}
For T6,
\begin{align*}
	&\textrm{T6}\\
	&=\ba{\hat{p}_1^{(-k)}\hat{\pi}_1^{(-k)} + \hat{p}_0^{(-k)}\hat{\pi}_0^{(-k)}}
	\frac{1}{\hat{p}^{{(-k)}}_1-\hat{p}^{{(-k)}}_0} \bb{\frac{\pi_1}{\hat{\pi}^{(-k)}_1}\frac{p_1}{\hat{p}^{{(-k)}}_1}-\frac{\pi_0}{\hat{\pi}^{(-k)}_0}\frac{p_0}{\hat{p}^{{(-k)}}_0}}\bb{e_{10}-\hat{e}^{(-k)}_{10}}
	\\
	&=\hat{\rho}^{{(-k)}}
	\frac{1}{\hat{p}^{{(-k)}}_1-\hat{p}^{{(-k)}}_0} 
	\bb{\frac{1}{\hat{\pi}^{(-k)}_1\hat{p}^{{(-k)}}_1\hat{\pi}^{(-k)}_0\hat{p}^{{(-k)}}_0}}
	\underbrace{\bb{\pi_1 p_1 \hat{\pi}^{(-k)}_0\hat{p}^{{(-k)}}_0 - \pi_0 p_0 \hat{\pi}^{(-k)}_1\hat{p}^{{(-k)}}_1}}_{(*)}
	\bb{e_{10}-\hat{e}^{(-k)}_{10}}
	\\ 
	& \textit{apply the decomposition that $ab-cd = \frac{1}{2}(a-c)(b+d) + \frac{1}{2}(a+c)(b-d)$ to $(*)$}
	\\ 
	&E^{(-k)}\ba{\textrm{-T6}} \lesssim
	\begin{aligned}
		&\norm{\hat{\pi}^{(-k)}_0 - \pi_0} \cdot \norm{\hat{e}^{(-k)}_{10} - e_{10}} 
		+
		\norm{\hat{\pi}^{(-k)}_1 - \pi_1} \cdot \norm{\hat{e}^{(-k)}_{10} - e_{10}} 
		\\
		&+\norm{\hat{p}^{(-k)}_0 - p_0} \cdot \norm{\hat{e}^{(-k)}_{10} - e_{10}} 
		+
		\norm{\hat{p}^{(-k)}_1 - p_1} \cdot \norm{\hat{e}^{(-k)}_{10} - e_{10}} 
	\end{aligned}. 	
\end{align*}

Therefore, we find the term (\ref{b-term2}) admits the following bias structure (using the abbreviated notation $r_f^{(-k)} = \norm{\hat{f}^{(-k)} - f}$) :
\begin{align*}
	&|\mc{I}_k|^{1/2} E^{(-k)} \ba{\widehat{M}^{(-k)} - M} = |\mc{I}_k|^{1/2} E^{(-k)} \ba{\textrm{T1} + \textrm{T2} + \textrm{T3} + \textrm{T4} - \textrm{T5} - \textrm{T6}} \\
	&\lesssim 	r_{\delta^*} \cdot r_{p_1}	+ r_{\delta^*} \cdot r_{p_0}	\\
	&+ r_{\delta^*} \cdot r_{\pi_1}	+ r_{\delta^*} \cdot r_{\pi_0}	\\
	&+ r_\phi \cdot r_{p_1} \\
	&+ r_w \cdot r_{\pi_1} + r_w \cdot r_{\pi_0}	\\
	&+ r_{e_{10}} \cdot r_{\pi_1} + r_{e_{10}} \cdot r_{\pi_0} \\
    &+ r_{e_{10}} \cdot r_{p_1} + r_{e_{10}} \cdot r_{p_0}
    \end{align*}
We further observe that the robustness property of the EIF is reflected in its bias structure; that is, the EIF remains unbiased when certain nuisance functions are evaluated at their true values (colored in red):
\begin{align*}
	&\mc{M}_1: \textit{models for }  p_z(X) \textit{ and }  \pi_z(X) \textit{ are evaluated at their true value, for } z\in\{0,1\};\\
	&|\mc{I}_k|^{1/2} E^{(-k)} \ba{\widehat{M}^{(-k)} - M}  \\
	&\lesssim 	r_{\delta^*} \cdot {\color{red}r_{p_1}}	+ r_{\delta^*} \cdot {\color{red}r_{p_0}}	\\
	&+ r_{\delta^*} \cdot {\color{red}r_{\pi_1}}	+ r_{\delta^*} \cdot {\color{red}r_{\pi_0}}	\\
	&+ r_\phi \cdot {\color{red}r_{p_1}} \\
	&+ r_w \cdot {\color{red}r_{\pi_1}} + r_w \cdot {\color{red}r_{\pi_0}}	\\
	&+ r_{e_{10}} \cdot {\color{red}r_{\pi_1}} + r_{e_{10}} \cdot {\color{red}r_{\pi_0}} \\
    &+ r_{e_{10}} \cdot {\color{red}r_{p_1}} + r_{e_{10}} \cdot {\color{red}r_{p_0}}
	\\ \\
	&\mc{M}_2: \textit{models for } \delta^*(X), e_{1z}(X) \textit{ and } w(X)\textit{ are evaluated at their true value, for } z\in\{0,1\};\\
	&|\mc{I}_k|^{1/2} E^{(-k)} \ba{\widehat{M}^{(-k)} - M}  \\
	&\lesssim 	{\color{red}r_{\delta^*}} \cdot r_{p_1}	+ {\color{red}r_{\delta^*}} \cdot r_{p_0}	\\
	&+ {\color{red}r_{\delta^*}} \cdot r_{\pi_1}	+ {\color{red}r_{\delta^*}} \cdot r_{\pi_0}	\\
	&+ {\color{red}r_\phi} \cdot r_{p_1} \\
	&+ {\color{red}r_w} \cdot r_{\pi_1} + {\color{red}r_w} \cdot r_{\pi_0}	\\
	&+ {\color{red}r_{e_{10}}} \cdot r_{\pi_1} + {\color{red}r_{e_{10}}} \cdot r_{\pi_0} \\
    &+ {\color{red}r_{e_{10}}} \cdot r_{p_1} + {\color{red}r_{e_{10}}} \cdot r_{p_0}.	
	\\ \\
	&\mc{M}_3: \textit{models for } \delta^*(X), e_{1z}(X) \textit{ and }  \pi_z(X) \textit{ are evaluated at their true value, for } z\in\{0,1\}\\
	&|\mc{I}_k|^{1/2} E^{(-k)} \ba{\widehat{M}^{(-k)} - M}  \\
	&\lesssim 	{\color{red}r_{\delta^*}} \cdot r_{p_1}	+ {\color{red}r_{\delta^*}} \cdot r_{p_0}	\\
	&+ {\color{red}r_{\delta^*}} \cdot {\color{red}r_{\pi_1}}	+ {\color{red}r_{\delta^*}} \cdot {\color{red}r_{\pi_0}}	\\
	&+ {\color{red}r_\phi} \cdot r_{p_1} \\
	&+ r_w \cdot {\color{red}r_{\pi_1}} + r_w \cdot {\color{red}r_{\pi_0}}	\\
	&+ {\color{red}r_{e_{10}}} \cdot {\color{red}r_{\pi_1}} + {\color{red}r_{e_{10}}} \cdot {\color{red}r_{\pi_0}} \\
    &+ {\color{red}r_{e_{10}}} \cdot r_{p_1} + r_{e_{10}} \cdot r_{p_0}.	
\end{align*}
\textbf{\textit{Asymptotic Property of (\ref{b-term3})}}

The expectation of the term (\ref{b-term3}) 
\begin{align*}
	&\mb{G}^{(-k)}_{\mc{I}_k} \ba{ \widehat{M}^{(-k)} - M} 
\end{align*}
has mean 0 conditional on $\mc{I}_k^{(-k)}$. And, the variance is 
\begin{align*}
	var^{(-k)}\bb{\mb{G}^{(-k)}_{\mc{I}_k} \ba{ \widehat{M} ^{(-k)} - M}}
	\leq 
	var^{(-k)}\bb{\ba{ \widehat{M}_i^{(-k)} - M} }
	\leq
	E^{(-k)}\bb{\ba{\widehat{M}_i^{(-k)} - M}^2}
\end{align*}
Then, it suffices to find the rate of $E^{(-k)}\bc{\bb{\widehat{M}_i^{(-k)} - M}^2}$. 
\\ \\
We first consider the expansion of $\hat{\theta}^{(-k)}$
\begin{align*}
	&\hat{\theta}^{(-k)} =
	\frac{\hat{\rho}^{(-k)}}{\hat{p}^{{(-k)}}_1-\hat{p}^{{(-k)}}_0} \frac{2Z-1}{ \hat{\pi}^{(-k)}_Z}  
    \bc{Y-A\hat{\delta}^{*(-k)}-Z\hat{\phi}^{(-k)} - \hat{w}^{(-k)}  - \frac{A}{\hat{p}^{{(-k)}}_Z} \bb{Y-Z\hat{\phi}^{(-k)} + \hat{e}^{(-k)}_{10}}
    }\\
	&=
	\ba{\hat{\rho}^{(-k)}-\rho}\frac{1}{\hat{p}^{{(-k)}}_1-\hat{p}^{{(-k)}}_0} \frac{2Z-1}{ \hat{\pi}^{(-k)}_Z}  
    \bc{
    Y-A\hat{\delta}^{*(-k)}-Z\hat{\phi}^{(-k)} - \hat{w}^{(-k)} - \frac{A}{\hat{p}^{{(-k)}}_Z} \bb{Y-Z\hat{\phi}^{(-k)} + \hat{e}^{(-k)}_{10}}
    }  \\
    &\quad   +
    \rho\bb{\frac{1}{\hat{p}^{{(-k)}}_1-\hat{p}^{{(-k)}}_0}-\frac{1}{p_1-p_0}} \frac{2Z-1}{ \hat{\pi}^{(-k)}_Z}  
    \bc{
    	Y-A\hat{\delta}^{*(-k)}-Z\hat{\phi}^{(-k)} - \hat{w}^{(-k)}- \frac{A}{\hat{p}^{{(-k)}}_Z} \bb{Y-Z\hat{\phi}^{(-k)} + \hat{e}^{(-k)}_{10}}
    }
    \\  
    &\quad   +
    \rho\frac{1}{p_1-p_0} \bb{\frac{1}{\hat{\pi}^{(-k)}_Z} - \frac{1}{\pi_Z}}  (2Z-1)
    \bc{
    	Y-A\hat{\delta}^{*(-k)}-Z\hat{\phi}^{(-k)} - \hat{w}^{(-k)} 
    	- \frac{A}{\hat{p}^{{(-k)}}_Z} \bb{Y-Z\hat{\phi}^{(-k)} + \hat{e}^{(-k)}_{10}}
    }   
    \\
    &\quad   +
    \rho\frac{1}{p_1-p_0}\frac{1}{\pi_Z}  (2Z-1)
    \bc{
    Y-A\ba{\hat{\delta}^{*(-k)}-\delta^*}-Z\ba{\hat{\phi}^{(-k)}-\phi} - \ba{\hat{w}^{(-k)}-w} 
    - A\bb{\frac{1}{p_Z}-\frac{1}{\hat{p}^{{(-k)}}_Z}} \bb{Y-Z\hat{\phi}^{(-k)} + \hat{e}^{(-k)}_{10}}
    }     
    \\
    &\quad   +
    \rho\frac{1}{p_1-p_0}\frac{1}{\pi_Z}  (2Z-1)
    \bc{
    	-A\delta^*-Z\phi - w - A \frac{1}{p_Z} \bb{Y- Z\ba{\hat{\phi}^{(-k)}-\phi} + \ba{\hat{e}^{(-k)}_{10}-e_{10}}}
    }
    \\     
    &\quad   +
    \rho\frac{1}{p_1-p_0}\frac{1}{\pi_Z}  (2Z-1)
    \bc{
    	-A \frac{1}{p_Z} \bb{ - Z\phi + e_{10}}
    } 
    \\
	&=
	\ba{\hat{\rho}^{(-k)}-\rho}\frac{1}{\hat{p}^{{(-k)}}_1-\hat{p}^{{(-k)}}_0} \frac{2Z-1}{ \hat{\pi}^{(-k)}_Z}  
    \bc{
    Y-A\hat{\delta}^{*(-k)}-Z\hat{\phi}^{(-k)} - \hat{w}^{(-k)} - \frac{A}{\hat{p}^{{(-k)}}_Z} \bb{Y-Z\hat{\phi}^{(-k)} + \hat{e}^{(-k)}_{10}}
    }  \\
    &\quad   +
    \rho\bb{\frac{1}{\hat{p}^{{(-k)}}_1-\hat{p}^{{(-k)}}_0}-\frac{1}{p_1-p_0}} \frac{2Z-1}{ \hat{\pi}^{(-k)}_Z}  
    \bc{
    	Y-A\hat{\delta}^{*(-k)}-Z\hat{\phi}^{(-k)} - \hat{w}^{(-k)}- \frac{A}{\hat{p}^{{(-k)}}_Z} \bb{Y-Z\hat{\phi}^{(-k)} + \hat{e}^{(-k)}_{10}}
    }
    \\  
    &\quad   +
    \rho\frac{1}{p_1-p_0} \bb{\frac{1}{\hat{\pi}^{(-k)}_Z} - \frac{1}{\pi_Z}}  (2Z-1)
    \bc{
    	-A\hat{\delta}^{*(-k)}-Z\hat{\phi}^{(-k)} - \hat{w}^{(-k)} 
    	- \frac{A}{\hat{p}^{{(-k)}}_Z} \bb{Y-Z\hat{\phi}^{(-k)} + \hat{e}^{(-k)}_{10}}
    }   
    \\
    &\quad   +
    \rho\frac{1}{p_1-p_0}\frac{1}{\pi_Z}  (2Z-1)
    \bc{
    -A\ba{\hat{\delta}^{*(-k)}-\delta^*}-Z\ba{\hat{\phi}^{(-k)}-\phi} - \ba{\hat{w}^{(-k)}-w} 
    - A\bb{\frac{1}{p_Z}-\frac{1}{\hat{p}^{{(-k)}}_Z}} \bb{Y-Z\hat{\phi}^{(-k)} + \hat{e}^{(-k)}_{10}}
    }     
    \\
    &\quad   +
    \rho\frac{1}{p_1-p_0}\frac{1}{\pi_Z}  (2Z-1)
    \bc{
    	 - A \frac{1}{p_Z} \bb{- Z\ba{\hat{\phi}^{(-k)}-\phi} + \ba{\hat{e}^{(-k)}_{10}-e_{10}}}
    }
    \\
    &\quad + \underbrace{\rho \frac{1}{p_1-p_0}\frac{2Z-1}{\pi_Z}\bb{Y - A\delta^* - Z\phi - w - \frac{A}{p_Z}\ba{Y-\phi+e_{10}}}}_{=\theta}
    .		
\end{align*}
Each element of $\ba{\widehat{M}_i^{(-k)} - M}$ is
\begin{align*}
	&\ba{\widehat{M}_i^{(-k)} - M}
	\\
	&
	=A\ba{\hat{\delta^*}^{*(-k)} - \delta^*} + 
	\hat{\theta}^{(-k)} -  \theta	
	\\
	&=
	A\ba{\hat{\delta^*}^{*(-k)} - \delta^*}  \\
	&\quad
	+
	\frac{\hat{\rho}^{(-k)}}{\hat{p}^{{(-k)}}_1-\hat{p}^{{(-k)}}_0} \frac{2Z-1}{ \hat{\pi}^{(-k)}_Z}  
    \bc{
    \begin{aligned}
    	&Y-A\hat{\delta}^{*(-k)}-Z\hat{\phi}^{(-k)} - \hat{w}^{(-k)} \\
    	&- \frac{A}{\hat{p}^{{(-k)}}_Z} \bb{Y-Z\hat{\phi}^{(-k)} + \hat{e}^{(-k)}_{10}}
    \end{aligned}
    }\\
    & \quad 
    -
    \frac{1}{p_1-p_0} \frac{2Z-1}{ \pi_Z}  
    \bc{
    \begin{aligned}
    	&Y-A\delta^*-Z\phi - w \\
    	&- \frac{A}{p_Z} \bb{Y - Z\phi + e_{10}}
    \end{aligned}
    }
	\\
	&=
	A\ba{\hat{\delta^*}^{*(-k)} - \delta^*} \tag{J1}\label{J1}  \\
	&\quad
	+
	\ba{\hat{\rho}^{(-k)}-\rho}\frac{1}{\hat{p}^{{(-k)}}_1-\hat{p}^{{(-k)}}_0} \frac{2Z-1}{ \hat{\pi}^{(-k)}_Z}  
    \bc{
    Y-A\hat{\delta}^{*(-k)}-Z\hat{\phi}^{(-k)} - \hat{w}^{(-k)} - \frac{A}{\hat{p}^{{(-k)}}_Z} \bb{Y-Z\hat{\phi}^{(-k)} + \hat{e}^{(-k)}_{10}}
    } \tag{J2}\label{J2}
    \\
    &\quad   +
    \rho\bb{\frac{1}{\hat{p}^{{(-k)}}_1-\hat{p}^{{(-k)}}_0}-\frac{1}{p_1-p_0}} \frac{2Z-1}{ \hat{\pi}^{(-k)}_Z}  
    \bc{
    	Y-A\hat{\delta}^{*(-k)}-Z\hat{\phi}^{(-k)} - \hat{w}^{(-k)}- \frac{A}{\hat{p}^{{(-k)}}_Z} \bb{Y-Z\hat{\phi}^{(-k)} + \hat{e}^{(-k)}_{10}}
    } \tag{J3}\label{J3}
    \\  
    &\quad   +
    \rho\frac{1}{p_1-p_0} \bb{\frac{1}{\hat{\pi}^{(-k)}_Z} - \frac{1}{\pi_Z}}  (2Z-1)
    \bc{
    	-A\hat{\delta}^{*(-k)}-Z\hat{\phi}^{(-k)} - \hat{w}^{(-k)} 
    	- \frac{A}{\hat{p}^{{(-k)}}_Z} \bb{Y-Z\hat{\phi}^{(-k)} + \hat{e}^{(-k)}_{10}}
    } \tag{J4}\label{J4}
    \\
    &\quad   +
    \rho\frac{1}{p_1-p_0}\frac{1}{\pi_Z}  (2Z-1)
    \bc{
    \begin{aligned}
    	&-A\ba{\hat{\delta}^{*(-k)}-\delta^*}-Z\ba{\hat{\phi}^{(-k)}-\phi} - \ba{\hat{w}^{(-k)}-w} \\
    	&- A\bb{\frac{1}{p_Z}-\frac{1}{\hat{p}^{{(-k)}}_Z}} \bb{Y-Z\hat{\phi}^{(-k)} + \hat{e}^{(-k)}_{10}}
    \end{aligned}
    } \tag{J5}\label{J5}  
    \\
    &\quad   +
    \rho\frac{1}{p_1-p_0}\frac{1}{\pi_Z}  (2Z-1)
    \bc{
    	 - A \frac{1}{p_Z} \bb{- Z\ba{\hat{\phi}^{(-k)}-\phi} + \ba{\hat{e}^{(-k)}_{10}-e_{10}}}
    } \tag{J6}\label{J6}
\end{align*}

For a finite number of random variables $\{W_1, \ldots, W_K\}$, there exists a constant $C$ satisfying 
\[
E\left\{\left(\sum_{j=1}^{K} W_j\right)^2\right\} \leq C \cdot E(W_j^2).
\]

Thus, it suffices to study the rate of the $\mc{L}(P)$-norm of each term, which are given in (\ref{J1})-(\ref{J5}) below.
\begin{align*}
	&(\text{\ref{J1}}): E\bc{ \bb{\rho\ba{\hat{\delta^*}^{*(-k)} - \delta^*}}^2} \lesssim \norm{\hat{\delta^*}^{*(-k)} - \delta^*}^2 = r_{\delta^*}^2;\\
	&(\text{\ref{J2}}): 
	E\bc{\ba{\hat{\rho}^{(-k)}-\rho}\frac{1}{\hat{p}^{{(-k)}}_1-\hat{p}^{{(-k)}}_0} \frac{2Z-1}{ \hat{\pi}^{(-k)}_Z}  
    \bc{
    Y-A\hat{\delta}^{*(-k)}-Z\hat{\phi}^{(-k)} - \hat{w}^{(-k)} - \frac{A}{\hat{p}^{{(-k)}}_Z} \bb{Y-Z\hat{\phi}^{(-k)} + \hat{e}^{(-k)}_{10}}
    } 
    } \\
	&\quad \quad \lesssim  \norm{\hat{p}_1^{(-k)}- p_1}^2  +  \norm{\hat{p}_0^{(-k)}- p_0}^2+\norm{\hat{\pi}_1^{(-k)}- \pi_1}^2  +  \norm{\hat{\pi}_0^{(-k)}- \pi_0}^2 = r_{p_1} + r_{p_0} + r_{\pi_1} + r_{\pi_0} \\
	&(\text{\ref{J3}}): 
	E\bc{
    \rho\bb{\frac{1}{\hat{p}^{{(-k)}}_1-\hat{p}^{{(-k)}}_0}-\frac{1}{p_1-p_0}} \frac{2Z-1}{ \hat{\pi}^{(-k)}_Z}  
    \bc{
    	Y-A\hat{\delta}^{*(-k)}-Z\hat{\phi}^{(-k)} - \hat{w}^{(-k)}- \frac{A}{\hat{p}^{{(-k)}}_Z} \bb{Y-Z\hat{\phi}^{(-k)} + \hat{e}^{(-k)}_{10}}
    }	
	}\\
	&\quad \quad  \lesssim \norm{\hat{p}_1^{(-k)}-p_1}^2 + \norm{\hat{p}_0^{(-k)}- p_0}^2 = r_{p_1} + r_{p_0}
	\\
	&(\text{\ref{J4}}): 
	E\bc{
    \rho\frac{1}{p_1-p_0} \bb{\frac{1}{\hat{\pi}^{(-k)}_Z} - \frac{1}{\pi_Z}}  (2Z-1)
    \bc{
    	-A\hat{\delta}^{*(-k)}-Z\hat{\phi}^{(-k)} - \hat{w}^{(-k)} 
    	- \frac{A}{\hat{p}^{{(-k)}}_Z} \bb{Y-Z\hat{\phi}^{(-k)} + \hat{e}^{(-k)}_{10}}
    }	
	}\\
	&\quad \quad  \lesssim \norm{\hat{\pi}_1^{(-k)}-\pi_1}^2 + \norm{\hat{\pi}_0^{(-k)}- \pi_0}^2 = r_{\pi_1} + r_{\pi_0}
	\\
    &(\text{\ref{J5}}): 
    E\bc{
    \rho\frac{1}{p_1-p_0}\frac{1}{\pi_Z}  (2Z-1)
    \bc{
    \begin{aligned}
        &-A\ba{\hat{\delta}^{*(-k)}-\delta^*}-Z\ba{\hat{\phi}^{(-k)}-\phi} - \ba{\hat{w}^{(-k)}-w} \\
        &- A\bb{\frac{1}{p_Z}-\frac{1}{\hat{p}^{{(-k)}}_Z}} \bb{Y-Z\hat{\phi}^{(-k)} + \hat{e}^{(-k)}_{10}}
    \end{aligned}
    }  
    }\\
    &\quad \quad  \lesssim \norm{\hat{\delta}^{*(-k)}-\delta^*}^2 + \norm{\hat{\phi}^{(-k)}-\phi}^2 + \norm{\hat{w}^{(-k)}-w}^2 + \norm{\hat{p}_1^{(-k)}-p_1}^2 + \norm{\hat{p}_0^{(-k)}- p_0}^2 \\
    &\quad \quad  = r_{\delta^*} + r_{\phi} + r_{w} + r_{\pi_1} + r_{\pi_0}
    \\	
    &(\text{\ref{J6}}): 
    E\bc{
    \rho\frac{1}{p_1-p_0}\frac{1}{\pi_Z}  (2Z-1)
    \bc{
         - A \frac{1}{p_Z} \bb{- Z\ba{\hat{\phi}^{(-k)}-\phi} + \ba{\hat{e}^{(-k)}_{10}-e_{10}}}
    } 
    }\\
    &\quad \quad  \lesssim \norm{\hat{\phi}^{(-k)}-\phi}^2  + \norm{\hat{e}_{10}^{(-k)}- e_{10}}^2 =  r_{\phi} + r_{e_{10}}    
\end{align*}

Combining these results, we find
\begin{align*}
	var^{(-k)}\bb{\mb{G}^{(-k)}_{\mc{I}_k} \ba{ \widehat{M}_i^{(-k)} - M}}
	\lesssim r_{\delta^*} + r_{\phi} + r_{w} + r_{p_1} + r_{p_0} + r_{\pi_1} + r_{\pi_0} + r_{e_{10}}.
\end{align*}
Under assumption \ref{as:consis} that $r_{p_z,N}^{(-k)}$, $r_{\pi_z,N}^{(-k)}$, $r_{e_{10},N}^{(-k)}$, $r_{w,N}^{(-k)}$ and $r_{\delta,N}^{(-k)}$ are $o_P(1)$ for $z\in \bb{0,1}$, $var^{(-k)}\bb{\mb{G}^{(-k)}_{\mc{I}_k} \ba{ \widehat{M}_i^{(-k)} - M}} = o_P(1)$, indicating $(\ref{b-term3})$ is $o_P(1)$ by Chebyshev's inequality (by showing it has $0$ expectation and shrinking variance).
\\

\textbf{\textit{Consistent Variance Estimation}}

Using the the shorthand notation $f = f(X)$ or $f=f(O)$, the consistent estimator of $\sigma^2$ proposed in \Cref{th:asym} is
\begin{align*}
        &\hat{\sigma}^2 = \frac{1}{K}\sum_{i=1}^K  \frac{1}{\mb{P}(A)}\mb{P}_{\mc{I}_k}
        \bc{
        \bb{
        \hat{\gamma}^{(-k)} - A\hat{\delta}^{*EIF}
        }^2
        }
        \\ 
        &where 
        \\
    &\hat{\gamma}^{(-k)}=    
    A\hat{\delta}^{*(-k)} + \theta^{(-k)}, \quad 
    \theta^{(-k)} = 
    \frac{\hat{\rho}^{(-k)}}{\hat{p}^{{(-k)}}_1-\hat{p}^{{(-k)}}_0} \frac{2Z-1}{ \hat{\pi}^{(-k)}_Z}  
    \bc{
    \begin{aligned}
    &Y-A\hat{\delta}^{*(-k)}-Z\hat{\phi}^{(-k)} - \hat{w}^{(-k)} \\
	&	- \frac{A}{\hat{p}^{{(-k)}}_Z} \bb{Y-Z\hat{\phi}^{(-k)} + \hat{e}^{(-k)}_{10}}
    \end{aligned}
    }.
    \end{align*}
which can also be written as
\begin{align*}
	&\hat{\sigma}^2 = K^{-1} \sum_{i=1}^K \hat{\sigma}^{2,(k)}, \quad 
	\hat{\sigma}^{2,(k)} = \mb{P}(A)^{-1}	\mb{P}_{\mc{I}_k}
	\bc{
	\bb{ A\hat{\delta}^{*(-k)}  + \hat{\theta}^{(-k)} - A\hat{\delta}^{*EIF}}^2}.
\end{align*}
Therefore, it suffices to show that \(\hat{\sigma}^{2,k} - \sigma^2 = o_P(1)\), which is represented as follows:
\begin{align*}
	&\sigma^2  = \bb{\mb{P}\ba{A}}^{-2} \mb{P}_{\mc{I}_k}
	\bc{
	\bb{
	{
	A\delta^* + \theta - A\delta^*_m
	}
	}^2
	}
	\\
	&\hat{\sigma}^{2,k} =\bb{\mb{P}\ba{A}}^{-2} \mb{P}_{\mc{I}_k}
	\bc{
	\bb{
	{
	A\delta^{*(-k)} + \theta^{(-k)} -A\hat{\delta}^{EIF}
	}
	}^2
	}
	\\
	&\hat{\sigma}^{2,k} - \sigma^2 
	=
	\bb{Pr(A=1)}^{-2} 
	\mb{P}_{\mc{I}_k}
	\bc{
	\bb{
	{
	A\delta^{*(-k)} + \theta^{(-k)} -A\hat{\delta}^{*EIF}
	}
	}^2
	-
	\bb{
	{
	A\delta^* + \theta -A\delta^*_m
	}
	}^2	
	} + o_P(1)
\end{align*}
The last equation holds by the law of large numbers. From some algebra, we find the term on the RHS of the last equation is
\begin{align*}
	&\mb{P}_{\mc{I}_k}
	\bc{
	\bb{
	{
	A\delta^{*(-k)} + \theta^{(-k)} - A\hat{\delta}^{*EIF}
	}
	}^2
	-
	\bb{
	{
	A\delta^* + \theta - A\delta^*_m
	}
	}^2	
	} \\
	&
	=\mb{P}_{\mc{I}_k}	
	\biggl[
	\bb{
	{
	A\delta^{*(-k)} + \theta^{(-k)} -A\hat{\delta}^{*EIF}
	}
	+
	{
	A\delta^* + \theta -A\delta^*_m
	}
	}	
	\\
	& \quad \quad \quad \cdot	
	\bb{
	A\delta^{*(-k)} + \theta^{(-k)} - A\hat{\delta}^{*EIF}
	-
	A\delta^* - \theta + A\delta^*_m
	}	
	\biggl]	\\
	&=\mb{P}_{\mc{I}_k}	
	\biggl[
	\bb{
	A\ba{\delta^{*(-k)} - \delta^* + \delta^*} + \bb{\theta^{(-k)}-\theta+\theta} - A\ba{\hat{\delta}^{*EIF} - \delta^*_m + \delta^*_m}
	+
	A\delta^* + \theta - A\delta^*_m
	}
	\\
	& \quad \quad \quad \cdot	
	\bb{
	A\ba{\delta^{*(-k)} - \delta^*} + \bb{\theta^{(-k)} -\theta} - A\ba{\hat{\delta}^{*EIF} - \delta^*_m}
	}	
	\biggl]	
	\\
	&=\mb{P}_{\mc{I}_k}	
	\biggl[
	\bb{
	A\ba{\delta^{*(-k)} - \delta^*} + \bb{\theta^{(-k)}-\theta} - A\ba{\hat{\delta}^{*EIF} - \delta^*_m}
	+
	2\bb{A\delta^* + \theta - A\delta^*_m}
	}
	\\
	& \quad \quad \quad \cdot	
	\bb{
	A\ba{\delta^{*(-k)} - \delta^*} + \bb{\theta^{(-k)} -\theta} - A\ba{\hat{\delta}^{*EIF} - \delta^*_m}
	}	
	\biggl]	
	\\
	&=\mb{P}_{\mc{I}_k}	
	\bc{
	\bb{
	A\ba{\delta^{*(-k)} - \delta^*} + \bb{\theta^{(-k)}-\theta} - A\ba{\hat{\delta}^{*EIF} - \delta^*_m}
	}^2
	} \label{DD1} \tag{D1}
	\\
	& \quad \quad \quad +	2 \mb{P}_{\mc{I}_k}	
	\bc{
	\bb{
	A\ba{\delta^{*(-k)} - \delta^*} + \bb{\theta^{(-k)} -\theta} -A\ba{\hat{\delta}^{*EIF} - \delta^*_m}
	}
	\bb{A\delta^* + \theta - A\delta^*_m}
	}	
	\label{DD2} \tag{D2}	
\end{align*}

Let $H ^{(-k)} = 	\bb{
	A\ba{\delta^{*(-k)} - \delta^*} + \bb{\theta^{(-k)} -\theta} +A\ba{\hat{\delta}^{*EIF} - \delta^*_m}
	}$. By the H\"older's inequality:
\begin{align*}
	&\bb{(\text{D1}) + (\text{D2})}^2
	\\
	&=
	\bc{
	\mb{P}_{\mc{I}_k} \bb{H^{(-k)}} + 2 \mb{P}_{\mc{I}_k} \bb{H^{(-k)}} \mb{P}_{\mc{I}_k} \bb{A\delta^* + \theta - A\delta^*_m}
	}^2
	\\
	&\lesssim \mb{P}_{\mc{I}_k}	 \bc{ \bb{H^{(-k)}}^2} + 2\mb{P}_{\mc{I}_k}	 \bc{ \bb{H^{(-k)}}^2} 
	\mb{P}_{\mc{I}_k}	 \bc{ \bb{A\delta^* + \theta - A\delta^*_m}^2}.
\end{align*}
Since $\mb{P}_{\mc{I}_k}	 \bc{ \bb{A\delta^* + \theta - A\delta^*_m}^2} = Pr(A=1) \sigma^2 + o_P(1) = O_p(1)$, $(\text{D1}) + (\text{D2})$ is $o_P(1)$ if 
$\mb{P}_{\mc{I}_k} \bc{ \bb{H^{(-k)}}^2}$ is $o_P(1)$. From some algebra, we find
\begin{align*}
	&\mb{P}_{\mc{I}_k} \bc{ \bb{H^{(-k)}}^2} \\
	&=\mb{P}_{\mc{I}_k} \bc{
	\bb{
	A\ba{\delta^{*(-k)} - \delta^*} + \bb{\theta^{(-k)} -\theta} +A\ba{\hat{\delta}^{*EIF} - \delta^*_m}
	}^2
	} \\
	&\leq 3\mb{P}_{\mc{I}_k}\bc{\bb{\theta^{(-k)} -\theta}^2} + 3\mb{P}_{\mc{I}_k}\bb{\ba{\delta^{*(-k)} - \delta^*}^2} + 3\mb{P}_{\mc{I}_k} \bb{\ba{\hat{\delta}^{*EIF} - \delta^*_m}^2}
	\\
	&= 3E^{(-k)}\bb{\ba{\theta^{(-k)} -\theta}^2} + o_P(1) + 3E^{(-k)}\bb{\ba{\delta^{*(-k)} - \delta^*}^2}+ o_P(1)+3E\bb{\ba{\hat{\delta}^{*EIF} - \delta^*_m}^2} + o_P(1) \\
	&=o_p(1) \quad \textit{(each estimate is consistent by assumption \eqref{as:consis})}
	.
\end{align*}
Therefore, the proposed estimator $\hat{\sigma}^2$ is consistent for $\sigma^2$
\begin{align*}
	&\hat{\sigma}^{2,k} - \sigma^2  =
	\bb{Pr(A=1)}^{-2} 
	\bb{(\text{D1}) + (\text{D2})} + o_P(1) = o_P(1).
\end{align*}

\end{appendices}
\end{document}